\documentclass[twocolumn,twoside]{IEEEtran}
\ifCLASSINFOpdf

\else

\fi

\ifCLASSOPTIONcompsoc
\usepackage[caption=false, font=normalsize, labelfont=sf, textfont=sf]{subfig}
\else
\usepackage[caption=false, font=normalsize]{subfig}
\fi
\usepackage{lipsum}% 

\usepackage{balance}
\usepackage{multicol}   % for equalize of last page columns
\usepackage{cite}
\usepackage{gensymb}
\usepackage{multirow}
\usepackage{graphics}  
\usepackage{epsfig} 
\usepackage{graphicx}
\usepackage{epstopdf}
\usepackage{textcomp}
\usepackage{amsmath}
\usepackage{mathtools}
\usepackage{filecontents}
\usepackage{lipsum,color}
\usepackage[dvipsnames]{xcolor}
\usepackage{amssymb}
\usepackage{float}
\usepackage{stfloats}
\usepackage{blindtext}
\begin{document}
	\title{Analytical Channel Model and Link Design Optimization for Ground-to-HAP Free-Space Optical Communications}
	
	\author{Hossein~Safi,~Akbar~Dargahi,~Julian~Cheng,~{\it Senior Member,~IEEE},~and~Majid~Safari,~{\it Member, IEEE}
		
		\thanks{H. Safi, and A. Dargahi are with the Department of Electrical Engineering, Shahid Beheshti University G. C., 1983963113, Tehran, Iran (e-mails: \{h$\_$safi, a-dargahi\}@sbu.ac.ir).  J. Cheng is with the School of Engineering, the University of British Columbia, V1V 1V7, Kelowna, BC Canada (e-mail: julian.cheng@ubc.ca). M. Safari is with the School of Engineering, the University of Edinburgh, EH8 9YL, Edinburgh, UK (e-mail: msafari@exseed.ed.ac.uk)}}
	%\thanks{Saeedeh~Parsaeefard ....}
	%}    
	
	% The paper headers
	%\markboth{IEEE~TRANSACRIONS~ON~WIRELESS~COMMUNICATIONS,~2018}
	%{Optical Channel Modeling for Ground-to-HAP Links over Turbulence Channels.}
	
	% make the title area
	\maketitle
	
	%%%%%%%%%%%%%%%%%%%%%%%%%%%%%%%%%%%%%%%%%%%%%%%%%%%%%%%%%%
	%%%%%%%%%%%%%%%%%%%%%%%%%%%%%%%%%%%%%%%%%%%%%%%%%%%%%%%%%%
	\begin{abstract}
		Integrating high altitude platforms (HAPs) and free space optical (FSO) communications is a promising solution to establish high data rate aerial links for
		the next generation wireless networks. However, practical limitations such as pointing errors and angle-of-arrival (AOA) fluctuations of the optical beam due to the orientation deviations of hovering HAPs  make it challenging to implement HAP-based FSO links. For a ground-to-HAP FSO link, tractable, closed-form statistical channel models are derived 
		in this paper to simplify optimal design of such systems. The proposed models include the combined effects of atmospheric turbulence regimes (i.e., log-normal and gamma-gamma), pointing error induced geometrical loss, pointing jitter variance caused by beam wander, detector aperture size, beam-width, and AOA fluctuations of the received optical beam. The analytical expressions are corroborated by
		performing Monte-Carlo simulations. Furthermore,  
		closed-form expressions for the outage probability  of the considered link under different turbulence regimes are derived. Detailed analysis is carried out to optimize the  transmitted laser beam and the field-of-view of the receiver for minimizing outage probability under different channel conditions. The obtained analytical results can be applied to finding the optimal parameter values and designing ground-to-HAP FSO links  without resorting to time-consuming simulations. 
		%%%%%%%%%%%%%%%%%%%%%%%%%%%%%%%%%%%%%%%%%%%%%%%%%%%%%%%%%%
		%%%%%%%%%%%%%%%%%%%%%%%%%%%%%%%%%%%%%%%%%%%%%%%%%%%%%%%%%%
	\end{abstract}
	\IEEEpeerreviewmaketitle
	%%%%%%%%%%%%%%%%%%%%%%%%%%%%%%%%%%%%%%%%%%%%%%%%%%%%%%%%%%%%
	%%%%%%%%%%%%%%%%%%%%%%%%%%%%%%%%%%%%%%%%%%%%%%%%%%%%%%%%%%%%
	\section{Introduction}
	%%%%%%%%%%%%%%%%%%%%%%%%%%%%%%%%%%%%%%%%%%%%%%%%%%%%%%%%%%%%
	%%%%%%%%%%%%%%%%
	
\begin{figure}[]
		\begin{center}
			\includegraphics[width=3. in ]{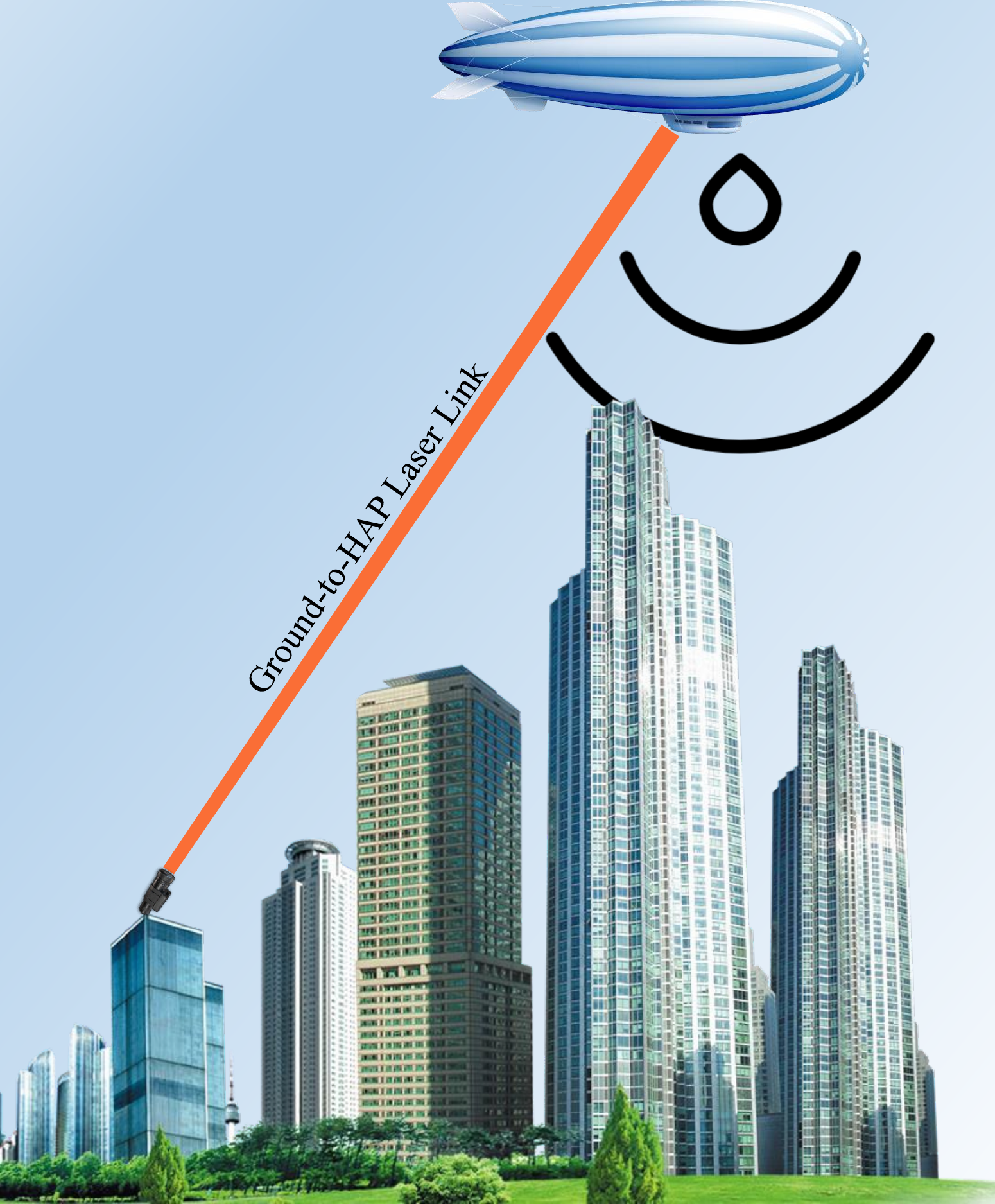}
			\caption{A typical illustration of a HAP-based FSO communication link. HAPs are usually placed in the stratosphere layer where the thin air is relatively calm and the wind speed is low, and thus, the link length is commonly larger than 17 km \cite{mozaffari2019tutorial}.}
			\label{typical_presentation}
		\end{center}
	\end{figure}

	%%%%%%%%%%%%%%%%%
	Recently, high altitude platforms (HAPs) have received considerable attention as a  promising candidate to extend the coverage of terrestrial networks by providing easy-to-deploy and cost-effective links \cite{mozaffari2019tutorial}. HAP systems
	are  preferable for providing broadband communications and wide-scale wireless
	coverage for large geographic areas \cite{karapantazis2005broadband}. In particular, HAPs can be used either as aerial relays to improve ubiquitous connectivity of terrestrial wireless systems, or as flying base stations (BSs) to provide reliable
	downlink and uplink communications for ground users \cite{mozaffari2017mobile}. Free space optical (FSO)-based front-haul/back-haul links are proposed as a promising approach for the next generation of wireless networks to confront the challenge of scarce radio spectrum resources  and to obtain high data rate transmission on the order of Gbps \cite{alzenad2018fso,dong2018edge}.
	To support FSO communications, as shown in Fig. \ref{typical_presentation}, HAPs can be considered as superior candidates. In particular, unique capabilities of HAPs, e.g., maneuverability and adaptive altitude adjustment enable them to effectively establish line-of-sight (LoS) communication links that are necessary for successful data transmission in an FSO link. As examples, Google’s Project Loon and Facebook’s Internet-delivery drone are the two recent projects  that combines  FSO communications with
	HAPs \cite{loonproject,faceebook}\footnote{\textcolor{black}{It is worth noting that, as proposed and implemented in \cite{5,6}, there exist spaceborne optical communication links (i.e.,  high data-rate bi-directional optical communications between Earth and geostationary Earth orbit (GEO), and low Earth orbit (LEO)) that employ adaptive optics (AO) to facilitate coupling the received optical signal into a single-mode fiber. In this regard, the AO system should be capable of coupling more than half the received signal into the single mode fiber. Due its technical complexity and implementation cost as well as it narrow scopes (which are mainly limited to deep space communications), in this paper, we do not consider these types of FSO communication systems that employ the AO subsystem in their links.}}.  
	
	%%%%%%%%%%%%%%%%
	%%%%%%%%%%%%%%%%%
	
	However, for successful implementation, FSO links suffer from practical limitations. First, FSO communication is sensitive to beam alignment from transmitter to receiver. Therefore, it is essential to perform accurate beam pointing at the transmitter side, and beam acquisition and tracking at the receiver side. With an array of photo-detectors located at the focal plane of the receiver, a spatial beam tracking method was proposed for a ground-to-drone FSO link to tackle the effect of hovering fluctuations of the receiver \cite{safi2019spatial}.   Moreover, there exist  accurate beam tracking methods based on mechanical or piezoelectric equipment,
	e.g., gimbals and retro reflectors, which are readily applicable for mobile FSO communications \cite{khan2017gps,kaadan2016spherical}. A recent comprehensive literature review \cite{kaymak2018survey}  discussed existing beam acquisition and tracking  mechanisms suitable for mobile FSO communications and categorized these mechanisms according to their working principles, use cases, as well as their advantages and disadvantages.
	
	Second, beam wander and scintillation due to inhomogeneity in
	temperature and pressure of the air can severely affect the link performance \cite{laserbook}. \textcolor{black}{To tackle the effect of atmospheric turbulence, different fading mitigation techniques such as novel sequence data detection algorithms \cite{song2014robust,dabiri2017glrt}, temporal and spatial diversity \cite{jaiswal2018investigation, mendenhall2007design,10.1117/12.273685}, aperture averaging, adaptive optics, and adaptive channel coding \cite{safi2019adaptive,walther2010air} have been proposed in the context of FSO. For instance, the authors in \cite{song2014robust} proposed  a Viterbi-type trellis-search sequence receiver based on the generalized likelihood ratio test principle that jointly detects the data sequence and estimates the unknown channel gain. The work in \cite{jaiswal2018investigation} considered a multiple-input multiple-output with optical space shift keying signaling scheme and employs transmit diversity to efficiently combat the turbulence effects. Moreover, the authors in \cite{mendenhall2007design} utilized a distributed photon-counting receiver array  as a cost-effective and adaptable alternative approach to traditional large, single-aperture receive elements. The work in \cite{10.1117/12.273685} used multiple transmitter to compensate scintillation fades.  Meanwhile, the authors in \cite{safi2019adaptive} presented adaptive coding and power transmission schemes to tackle the effect of atmospheric turbulence in a practical FSO communication system, and the use of forward error correction codes and interleaving is proposed in \cite{walther2010air} to combat power
		fluctuations from turbulence-induced scintillation.} \textcolor{black}{The turbulence effect can also be reduced by adaptive optics \cite{AOT,AOT1,AOT2}. By
		this technique, the distortion induced in the wave-front by the
		atmospheric turbulence is reduced through the use of wavefront sensors and deformable mirrors \cite{khalighi2014,kaushal2017optical}. However, the application of this technique is limited and it
		does not seem to be of interest in commercial FSO systems due to its high and unjustified implementation complexity and
		cost \cite{khalighi2014}. In addition, its effectiveness to compensate the effects of atmospheric turbulence is
		practically limited to relatively short-range links \cite{AOT3,kaushal2017optical}.}
	
	Third,
	mounting the optical receiver on a HAP station can cause the angle-of-arrival (AOA) fluctuations due to orientation 
	deviations of the receiver, which in turn induce signal-to-noise ratio
	(SNR) fluctuations and significantly degrade the reliability of the system \cite{huang2017free}. Indeed, such degradation factors are distance-dependent and their effects are significant on long-range FSO links, which is usually the case for HAP-based FSO systems.
	Therefore, the impairments caused by these factors  should be taken into account when evaluating the performance of such communication systems.
	%
	% Therefore, it is essential to have a detailed channel model to assess the benefit of such links,
	%\footnote{HAPs are usually placed in the stratosphere layer where the thin air is relatively calm and the wind speed is low, and thus, the link length is commonly larger than 17 km \cite{mozaffari2019tutorial}.}. 
	%
	% Another serious challenge raised by mounting the optical receiver on a HAP station is the angle-of-arrival (AOA) fluctuations due to orientation 
	%deviations of the receiver. In particular, AOA fluctuations of optical beam at the receiver aperture induce signal-to-noise ratio
	%(SNR) fluctuations that can significantly degrade the reliability of the system \cite{huang2017free}. 
	
	To assess the benefits of a ground-to-HAP FSO link, the communication channel should be distinctively characterized in terms of the receiver random vibrations due to hovering fluctuations and optical beam propagation characteristics in the atmosphere. Although, there has been a surge of recent works on drone based FSO communications \cite{alzenad2018fso,dong2018edge,fawaz2018uav,yang2019performance,li201780,mai2019beam,li2018investigation}, these prior works all assumed  stable drones and did not address the presence of AOA fluctuations and position vibrations. Recently, studies have been reported on the effects of random fluctuations in the aperture position and orientation as well as atmospheric turbulence loss and attenuation. For examples, a multi rotor drone-based FSO link was modeled to take into consideration of the effects due to position and AOA fluctuations \cite{dabiri2018channel}. However, the proposed model is quite complex and not so tractable for further research
	investigations. More recently, a simpler and tractable channel model for the considered system model in \cite{dabiri2018channel} are proposed in \cite{dabiri2019tractable} over log-normal atmospheric turbulence environment. However, the authors in \cite{dabiri2018channel, dabiri2019tractable} ignored the effects of the side-lobes of optical Airy pattern at the receiver which result in an outage probability floor. Meanwhile, based on the assumption of non-orthogonal incident beam to the photo detector (PD) plane, a statistical model was proposed for the geometrical and misalignment losses of the FSO channel
	for unmanned aerial vehicles (UAVs)-based networks  \cite{najafi2019statistical}, where the background noise was assumed as the dominant noise source at the PD. However, in this noise regime, the receiver field-of-view (FOV) was not optimized to mitigate the effects of background noise and orientation deviations of the UAV. In the aforementioned works, it is commonly assumed that the transceiver has the same altitude (i.e., not a slant path). In addition, the effect of beam wander is typically neglected due to short link length (low-altitude assumptions), and also the pointing error geometrical loss model is developed based on Gaussian beam profile. However, this assumptions may not hold for a HAP-based FSO system.
	%\\
	%\\
	%\\
	%\\
	%\\
	%In this noise regime, however, the importance of optimizing the receiver field-of-view (FOV)  to tackle the effect of background noise as well as to alleviate the effects of the orientation deviations of the UAV has not been discussed in \cite{najafi2019statistical}.  Nevertheless, these prior art on channel modeling assumed that the transceivers are in the same altitude (i.e., not a slant path). In addition, the link length is short enough to neglect the effect of beam wander and also develop the pointing error geometrical loss model based on Gaussian beam profile. However, this approach does not properly capture the channel model for a HAP-based FSO system. 
	Because the transceivers in such systems do not have the same altitude, and also the link distance is long. As a result, plane wave and spherical wave models are more accurate optical wave models than the Gaussian beam profile  \cite{ laserbook,gagliardi1995optical} for presenting the characteristic of beam profile at the receiver.
	
	In this paper, we drive analytical channel models for  ground-to-HAP FSO links by taking into account the effects of atmospheric attenuation and turbulence (both log-normal (LN) and gamma-gamma (GG) turbulence models), pointing error induced geometrical loss, and the effects of hovering fluctuations of the HAP, i.e.,  position vibrations of the optical receiver as well as the AOA fluctuations of the received optical beam. We first consider optical beam profile at the receiver for a long-range FSO link, and propose a new statistical model for pointing error induced geometrical loss. This model incorporates the position vibrations of the receiver, pointing jitter variance caused by
	beam wander, detector aperture size, and received optical beam-width. Using the developed pointing error model, we drive closed-form expressions for the channel model of the considered link under different turbulence regimes, i.e., weak and moderate to strong atmospheric turbulence conditions. For FSO systems, the  coherence time of the communication channel  is on the order of 1-100 \textrm{msec} which is significantly greater than the typical nanosecond bit duration (or equally Gbps transmission rate) \cite{zhu2002free}. Therefore, for such slow fading channels, outage probability, which is the probability of the event when  the instantaneous SNR falls bellow a certain threshold, is the most relevant metric to evaluate system performance \cite{khalighi2014}. Accordingly, we derive 
	closed-form expressions for the outage probability of the considered link for both LN and GG atmospheric turbulence models. Moreover, we provide a detailed analysis for optimizing the  transmitted laser beam by tuning divergence angle of the transmitter and the FOV of the receiver to achieve minimum outage probability under different channel conditions. In particular, we show that optimizing the beam-width of the transmitter calls for balancing a tradeoff between the amounts of pointing error and effective transmitter gain. Furthermore, when optimizing the receiver FOV,  a compromise is required between the amount
	of undesired background power and mitigation of beam position deviation, which is due to hovering fluctuations.  Simulation results are provided to validate the derived analytical expressions of the channel model and outage probability. 
	Thus, from the developed analytical expressions for channel modeling and outage probability, performance evaluation of the ground-to-HAP FSO links can be carried out without resorting to time-consuming simulations.
	
	The rest of the paper is organized as follows. Section II presents the system model. In Section III, we introduce the statistical model for pointing error induced geometrical loss and derive  channel distribution functions and outage probability of the considered link. In Section IV, numerical results are provided to demonstrate the need for optimizing receiver divergence angle and transmitter FOV under different channel conditions. Finally, we conclude the paper in Section V.
	%%%%%%%%%%%%%%%%%%%%%%%%%%%%%%%%%%%%%%%%%%%%%%%%%%%%%%%%%%%%
	%%%%%%%%%%%%%%%%%%%%%%%%%%%%%%%%
	%\subsection{Background}
	%%%%%%%%%%%%%%%%%%%%%%%%%%%%%%%%
	%\IEEEPARstart{}{}

	%%%%%%%%%%%%%%%%%%%%%%%%%%%%%%%%%%%%%%%%%%%%%%%%%%%%%%%%%%%%
	%%%%%%%%%%%%%%%%%%%%%%%%%%%%%%%%%%%%%%%%%%%%%%%%%%%%%%%%%%%%
	\section{System Model}
	%%%%%%%%%%%%%%%%%%%%%%%%%%%%%%%%%%%%%%%%%%%%%%%%%%%%%%%%%%%%
	%%%%%%%%%%%%%%%%%%%%%%%%%%%%%%%%%%%%%%%%%%%%%%%%%%%%%%%%%%%%
	%%%%%%%%%%%%%%%%%%%%%%%
	%%%%%%%%%%%%%%%%%%%%%%%
	\begin{figure}[!]
		\begin{center}
			\includegraphics[width=3.5 in ]{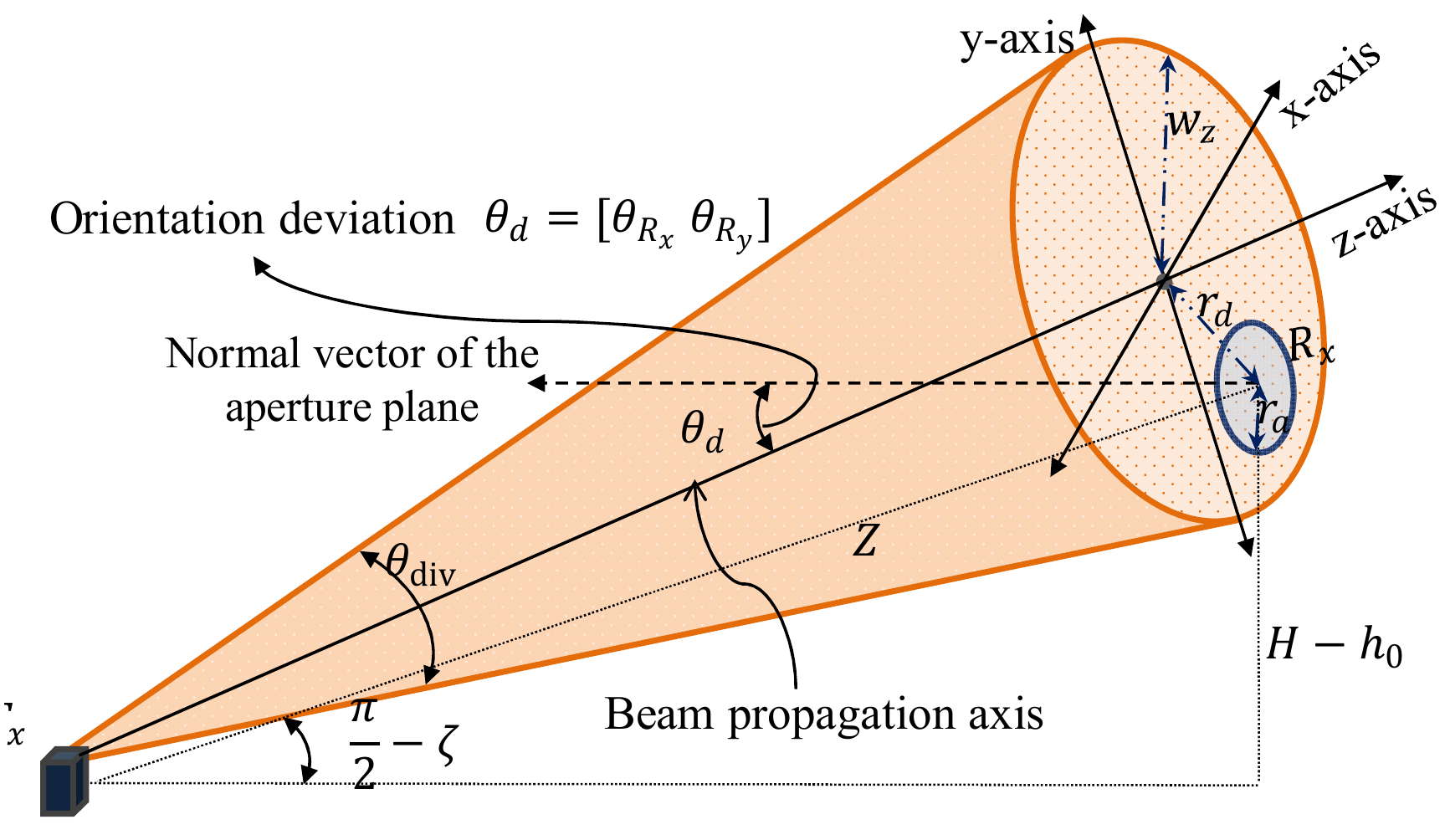}
			\caption{\textcolor{black}{The schematic of the considered optical uplink. Link parameters are defined as follows. $Z$ is the link length, $H$ is the HAP altitude, $h_0$ is the transmitter altitude, $\theta_\textrm{div}$ is the divergence angle of the optical transmitter,   $\zeta$ is the HAP zenith angle, $r_d$ denotes the separation distance between the center of optical beam footprint and the center of the receiver aperture, $r_a$ is the aperture radius,  and $w_Z$ is the radius of the received optical beam at distance $Z$. Also,  the orientation deviations of the HAP are indicated by $\theta_{Rx}$ and $\theta_{Ry}$ in $x-z$ plane and $y-z$ plane, respectively.}}
			\label{2}
		\end{center}
		%\vspace*{-6pt}
	\end{figure}
	%%%%%%%%%%%%%%%%%%%%%%%
	%%%%%%%%%%%%%%%%%%%%%%%
	The schematic of the considered optical uplink is shown in Fig. \ref{2} where a ground station transmits optical signals towards an aerial platform. We assume that the aerial-based receiver is hovering at a distance $Z$ from the ground transmitter. Also, the mean position of the receiver is  $H_{R}=\left(0,0,0\right)$, and its location is known to the transmitter. Regarding $H_{R}$, the instantaneous position of the HAP is indicated by $H_{d}=\left(d_x,d_y,d_z\right)$, where the independent random variables (RVs) $d_x$, $d_y$, and $d_z$ denote random deviations along the axes of coordinates. Since the link length $Z$ is much larger than the variance of  RV $d_z$, compared to the RVs $d_x$ and $d_y$, one can reasonably assume to neglect the amount of vibration along the $z$-axis.  In practice, due to the effects of hovering, the aperture position and orientation can deviate away from their means, causing fluctuations in the AOA of the received optical beam. Accordingly, the independent RVs of orientation deviations of the HAP are denoted by $\theta_{Rx}$ and $\theta_{Ry}$ in $x-z$ plane and $y-z$ plane, respectively. We also assume that the instantaneous position and the orientation deviations of the HAP node are Gaussian distributed \cite{dabiri2018channel}, i.e., $d_x$, $d_y$ $\sim \mathcal{N}(0,\sigma_d^2)$, and $\theta_{Rx}$,  $\theta_{Ry}$ $\sim \mathcal{N}(0,\sigma_0^2)$. Moreover, at the receiver aperture, the random displacements of the centroid of the
	propagated beam due to beam wander effects along the $x$ and $y$ coordinates, namely $b_x$ and $b_y$, are approximately Gaussian
	distributed with mean zero and variance \cite{li2018investigation}
	\begin{align}
	\label{variance_beam_wander}
	\sigma_{b}^2 = 2.07
	\int_{h_0}^{H} C_n^2\left( l \right)\left(Z-l\right)^2w_h^{-\frac{1}{3}}dl
	\end{align}
	where $H$, $h_0$, and $w_l$ are respectively the HAP altitude, the transmitter altitude, and the beam-width at distance $l$. The propagation distance $Z$ is given by $(H-h_0)\sec(\zeta)$, where $\zeta$ is the HAP zenith angle\setcounter{footnote}{0}\footnote{Zenith angle is the angle between two straight lines from the ground transmitter to the HAP and the zenith point.}. Moreover,
	In \eqref{variance_beam_wander}, $C_n^2\left( l \right)$ is  the refractive-index structure parameter that describes the varying strength of optical turbulence as a function of altitudes $l$, and it can be obtained as \cite[p.481]{laserbook}
	\begin{align}
	\label{rayrov}
	C_n^2\left( l \right) =~& 0.00594\left(V_w/27 \right)^2\left(10^{-5}l \right)^{10} e^{-l/1000} \nonumber \\
	&+ 2.7\times10^{-16} e^{-l/1500}+S_te^{-l/100}
	\end{align}
	where $V_w$ is the root-mean-square (rms) wind speed in meter per second (m/s), and $S_t$ is the nominal value of $C_n^2(0)$ at the ground in $\text{m}^{-2/3}$.

	The incoming optical beam is focused onto the PD through the lens. Due to the scattered sunlight, the PD inevitably collects undesired background light. Assuming intensity modulation at the transmitter and direct detection at the receiver, the PD converts the optical signal to an electrical current. For the $i$th symbol interval, the output photo-current can be obtained as \cite{dabiri2017fso}
	\begin{align}
	\label{sg1}
	r[i] = \eta\, h\, s[i]  +  n[i]
	\end{align}
	where $\eta$, $h$, $s[i]$, and $n[i]$ are, respectively, the PD responsibility, the channel coefficient, the transmitted symbol with average optical power $P_t$, and  the signal-independent zero-mean Gaussian noise with variance $\sigma_{n}^2$. Under the assumption of background noise as the dominant noise source at the receiver, the noise variance $\sigma_{n}^2$ can be expressed as \cite{laserbook}
	\begin{align}
	\label{noise_variance}
	\sigma_{n}^2 = 2eB_e\eta P_b
	\end{align}
	where $e$ is the electron charge, $B_e$ denotes the bandwidth of the PD, and $P_b$ is the power of background light. Also, we have  \cite{laserbook}
	\begin{align}
	\label{background_power}
	P_b =N_bB_o\Omega_{\textrm{FOV}}A_r
	\end{align}
	where $N_b$ denotes the spectral radiance of the background radiations at wavelength $\lambda$\footnote{\textcolor{black}{We note that, for the considered HAP-based FSO link, the choice of wavelength strongly depends on atmospheric
			effects, attenuation and background noise power, the availability of transmitter and receiver components, eye
			safety regulations, and cost. Indeed, choosing the operating
			wavelength requires balancing a tradeoff between the receiver
			sensitivity and pointing bias due to thermal variations
			across the Earth's surface. Thus, longer wavelengths are
			preferred as they make reduction in solar scattering as well as solar background from the Earth surface \cite{kaushal2017optical}.}}, $B_o$ stands for the bandwidth of the optical filter at the receiver, and $A_r$ denotes the aperture area. \textcolor{black}{ It is worth mentioning that, the spectral radiance of the sky, $N_b(\lambda)$, is measured for a rectangular portion of the horizon sky under various weather conditions and at
		different solar positions \cite{NB2,NB3,NB4}. Accordingly, it is a function of wavelength,  zenith angle, and azimuth angle between the telescope, the
		target and the Sun, and  is commonly derived by sequentially
		measuring a set of directions in the sky (i.e., sky scanning). Moreover, the sky radiance for different parameters has been computed using MODTRAN, a worldwide computer program designed to model atmospheric propagation of electromagnetic radiation, and the results are provided in \cite{NB1}. For instance, from the result of \cite{NB1} and for $\lambda = 1500$ nm, solar zenith angle
		of 45$^{\circ}$, and azimuthal angle of 0$^{\circ}$, the spectral radiance is equal to $10^{-3}$ W/$\text{cm}^2$-m-srad.}

	Also,  $\Omega_{\textrm{FOV}}$ in \eqref{background_power} is the receiver FOV, i.e., the solid angle through which the receiver can capture the transmitted laser beam, and it is obtained in the spherical coordinate system as 
	\begin{align}
	\label{FOV}
	\Omega_{\textrm{FOV}} = 2\pi\big(1-\cos(\theta_{\textrm{FOV}}/2)\big)
	\end{align}
	where $\theta_{\textrm{FOV}} = 2\arctan(\frac{r_p}{f_c})$, and where $r_p$ and $f_c$ are, respectively, the radius of the circular PD and focal length of the aperture. Furthermore, the instantaneous electrical signal-to-noise ratio (SNR) is obtained as \cite{laserbook}
	\begin{align}
	\label{SNR}
	\gamma = \dfrac{\eta^2P_t^2h^2}{\sigma_n^2}.
	\end{align}
	\textcolor{black}{Since in FSO systems the coherence time of all different channel variables are long relative to the bit duration (i.e., slow fading channel), when no fading-mitigation technique such as aperture
		averaging, diversity, or adaptive optics is employed, outage probability becomes more meaningful for evaluating the systems performance \cite{pointing2007}.} 
	Accordingly, the outage probability is defined as the probability that the instantaneous SNR is less than a threshold $\gamma_\textrm{th}$, and it can be written as
	\begin{align}
	\label{outage}
	\mathbb{P}_{\textrm{out}} = \int_{0}^{\gamma_\textrm{th}} f_{\gamma}(\gamma) d\gamma
	\end{align}
	where $f_{\gamma}(\gamma)$ is the probability density function (PDF) of $\gamma$. Since $\gamma$ is a monotonically increasing function of $h$, the outage probability can be also obtained as
	\begin{align}
	\label{outage2}
	\mathbb{P}_{\textrm{out}} = \int_{0}^{h_\textrm{th}} f_{h}(h) dh
	\end{align}
	where $h_\textrm{th} = \frac{\sqrt{\gamma_\textrm{th}\sigma_n^2}}{\eta P_t}$, and $f_h(h)$ is the PDF of $h$.  
	\section{ Channel Modeling}
	%%%%%%%%%%%%%%%%%%%%%%%%%%%%%%%%%%%%%%%%%%%%%%%
	%%%%%%%%%%%%%%%%%%%%%%%%%%%%%%%%%%%%%%%%%%%%%%%
	%%%%%%%%%%%%%%%%%%%%%%%%%%%%%%%%%%%%%%%%%%%%%%%
	%%%%%%%%%%%%%%%%%%%%%%%%%%%%%%%%%%%%%%%%%%%%%%%
	Four channel parameters are incorporated into $h$, i.e., 
	\begin{align}
	\label{hj5}
	h = \underbrace{h_{al} h_{at} h_{pl}}_{h_{ag}} h_{af}
	\end{align}
	where $h_{al}$, $h_{at}$, $h_{pl}$, and $h_{af}$ stand for the attenuation loss, the atmospheric turbulence,
	the effective pointing error induced geometrical loss, and the link interruption due to the AOA fluctuations at the receiver, respectively. In the sequel, the link interruption parameter $h_{af}$ takes two discrete values ``1'' or ``0'' to indicate the presence or absence of received beam in the receiver FOV. 
	%More details about these parameters are provided in the sequel.	
	%%%%%%%%%%%%%%%%%%%%%%%
	%%%%%%%%%%%%%%%%%%%%%%%
	\subsection{Attenuation loss and Atmospheric Turbulence}
	%%%%%%%%%%%%%%%%%%%%%%%
	%%%%%%%%%%%%%%%%%%%%%%%
	For an optical link with length $Z$,  the attenuation loss is represented by the  Beers-Lambert law as $h_{al}=\exp\left(-Z\xi\right)$, where $\xi$ is the attenuation coefficient related to the visibility \cite{al2004fog}.
	%Considering the atmospheric turbulence induced fading, we adopt the simple log-normal model, which is appropriate for relatively weak turbulence conditions \cite{ghassemlooy2012optical}. This can be justified by considering relatively short link ranges and assuming sufficient aperture averaging \cite{khalighi2009fading}. The distribution of $h_{at}$ is then given by
	%\begin{align}
	%	\label{ch1}
	%	f_{h_{at}}(h_{at})=\frac{1}{2h_{at} \sigma_{{\rm Ln}h_{at}}\sqrt{2\pi}} \exp\left( -\frac{\left(\ln(h_{at})-2\mu_{{\rm Ln}h_{at}}\right)^2}{8\sigma^2_{{\rm Ln}h_{at}}}\right).
	%\end{align}
	%where $\mu_{{\rm Ln}h_{at}}$ and $\sigma^2_{{\rm Ln}h_{at}}$  denote the mean and variance of log-irradiance, respectively, where $\sigma^2_{{\rm Ln}h_{at}}\simeq \sigma^2_{R}/4$ with $\sigma_R^2$ being the Rytov variance \cite{khalighi2009fading}. Normalizing the fading coefficient, i.e., setting $\mathbb{E}[h_{at}]=1$ (with $\mathbb{E}[.]$ denoting the expected value) we have $\mu_{{\rm Ln}h_{at}}=-\sigma^2_{{\rm Ln}h_{at}}$.
	To model the atmospheric turbulence induced fading, we consider both LN and GG  atmospheric turbulence models. Accordingly, the LN model is appropriate for
	weak turbulence conditions whereas the GG model is a suitable statistical model for moderate to strong atmospheric turbulence conditions\cite{laserbook}. The PDF of $h_{at}$ based on the LN model is obtained as 
	\begin{eqnarray}
	\label{fg1g-LN}
	f_{\rm {LN}}(h_{at})=  \dfrac{1}{2h_{at}\sqrt{2\pi\sigma_{Bu}^2}}\exp\left(-\frac{(\ln h_{at} + 2\sigma_{Bu}^2)^2}{8\sigma_{Bu}^2}\right)
	\end{eqnarray}
	where $\sigma_{Bu}^2$ is the Rytov variance. Note that, the transceivers in our setup are not located at the same height (i.e., a slant path), $\sigma_{Bu}^2$ is obtained as\footnote{\textcolor{black}{The Earth's atmosphere is composed of several distinct layers extends to approximately 700 km above the Earth’s surface with the heaviest concentration of particles in the first 40 km above the surface (also known as the free atmosphere). In our work we rely on the proposed general profile model, also known as Hufnagle-Valley model, in \cite{laserbook}. The H-V model is well suited for gounod-to-air links and covers the link range of  several tens of kilometers (which is the case for the considered HAP-based FSO communications in this work) \cite{H-V,H-V2,H-V3}. It is worth mentioning that, standard atmospheric spectral models are based on isotropic conditions throughout
			the free atmosphere. However, there exists evidence that reveals the turbulence above the free atmosphere is nonisotropic. For the communication links longer than 40 km, this will affect the computations of outage probability and the other link parameters.}} \cite{laserbook}
	\begin{align}
	\label{SI}
	\sigma_{Bu}^2 =& 2.25\left(\frac{2\pi}{\lambda} \right)^\frac{7}{6}
	\left(H-h_0\right)^\frac{5}{6}\sec(\zeta)^\frac{11}{6}
	\\ \nonumber
	&\times\int_{h_0}^{H} C_n^2\left( l \right)\left(1-\frac{l-h_0}{H-h_0}  \right)^\frac{5}{6}\left(\frac{l-h_0}{H-h_0}  \right)^\frac{5}{6}dl.
	\end{align}
	Moreover, the PDF of $h_{at}$ according
	to the GG model is obtained as 
	\begin{eqnarray}
	\label{fg1g}
	f_{\rm {GG}}(h_{at})=  \frac{2(\alpha\beta)^{\frac{\alpha+\beta}{2}}}{\Gamma(\alpha)\Gamma(\beta)} 
	h_{at}^{\frac{\alpha+\beta}{2}-1}   
	K_{\alpha-\beta}(2\sqrt{\alpha\beta h_{at}}) 
	\end{eqnarray}
	where $ \Gamma (\cdot) $ is the Gamma function and $ K_{\nu}(\cdot) $ is the modified Bessel function of the second kind of order $\nu$. Also, parameters $ \alpha $ and $ \beta $ denote the effective numbers of large-scale and small-scale eddies, respectively, and they can be represented as \cite{laserbook}
	\begin{align}
	\label{alpbet}
	\alpha &= \left[\exp\left( 0.49\sigma_{Bu}^2\Big/\left(1+0.56\sigma_{Bu}^{12/6}\right)^{7/6}\right)-1\right]^{-1}, \nonumber \\
	\beta  &= \left[\exp\left( 0.51\sigma_{Bu}^2\Big/\left(1+0.69\sigma_{Bu}^{12/6}\right)^{5/6}\right)-1\right]^{-1}.
	\end{align}
	\subsection{Effective Pointing Error}
	%%%%%%%%%%%%%%%%%%%%%%%
	%%%%%%%%%%%%%%%%%%%%%%%
	%The spherical wave model is sometimes used for a small-aperture source or a source with a large divergence angle.
	
	%Separate treatment is given for plane waves, spherical waves, and Gaussian-beam waves, and results are valid under all conditions of irradiance fluctuations.
	
	%Spherical wave are assumed when the source excitation for FSO communication can be considered as a point source.
	
	%divergent spherical wave

	%The field intensity of an spherical wave at distance $Z$ for an isotropic source is \cite[eqs. (1.2.5) and (1.2.6)]{gagliardi1995optical}
	Considering a Gaussian optical beam at the transmitter, the normalized spatial distribution of the
	transmitted intensity at distance $Z$ is given by \cite{saleh2019fundamentals}
	\begin{align}
	\label{int_Gaussian}
	I(\rho;Z) = \frac{2}{\pi\omega_Z^2}\exp\left(-\frac{2||\boldsymbol{\rho}||^2}{\omega_Z^2}\right)
	\end{align}
	where $\boldsymbol{\rho}$ is the radial vector form the center of the optical beam. Also in \eqref{int_Gaussian}, $\omega_Z$ is the beam waist at distance $Z$ and is approximately obtained as \cite{pointing2007}
	\begin{align}
	\label{w_z}
	\omega_Z 	\approx w_0{\Bigg[1 + \epsilon\bigg(\frac{\lambda Z}{\pi \omega_0^2}\bigg)^2\Bigg]}^{0.5}
	\end{align} 
	where $\omega_0$ is the beam waist at $Z=0$, and it is a parameter that can be tuned at the optical transmitter by employing different diameters for the aperture (or equivalently different divergence angles). More precisely, for a transmitter aperture of diameter $D$, $w_0$ is equal to $\frac{D}{\sqrt{2}\pi}$ \cite{siegman1986lasers}. Also in \eqref{w_z}, $\epsilon = \big(1+2\omega_0^2/\rho_0^2\big)$, and $\rho_0 = \int_{h_0}^{H} \big(0.55C_n^2\left( l \right)\left( \frac{2\pi}{\lambda} \right)^2l\big)^{-\frac{3}{5}}dl$ is the coherence length. Let $\boldsymbol{r_d}$ denote the radial beam displacement vector, i.e., the separation distance between the center of optical beam footprint and the center of the receiver aperture due to pointing errors. 
	Therefore, for a circular receiver aperture with radius $r_a$
	and the collecting area $A_r = \pi r_a^2$ which is aligned at an angle $\theta_d$ with respect to the arriving beam direction, pointing error loss due to geometrical spread can be expressed as
	\begin{align}
	\label{pointing_loss}
	h_{pl|\theta_d} = \int_{A_r}I(\boldsymbol{\rho} - \boldsymbol{r_d};Z)\cos(\theta_d)d\boldsymbol{\rho}.
	\end{align}
	
	In the considered link, the receiver is located at a distance of $Z$ from the optical transmitter where $\omega_Z \gg r_a$. In this regime, the overall phase of the wave becomes constant and phase difference between different parts of the optical wavefront at the receiver can be ignored. Hence, the  optical field at the receiver lens maintains locally the plane wave nature \cite{gagliardi1995optical,saleh2019fundamentals}. Thus, when the aperture area is much smaller than the beam waist, we can reasonably assume that the received optical intensity over the aperture area is constant, and we rewrite \eqref{pointing_loss} as 
	\begin{align}
	\label{pointing_loss_app}
	h_{pl|\theta_d}\simeq  2\left(\frac{r_a}{\omega_Z^2}\right)^2\exp\left(-\frac{2||\boldsymbol{r_d}||^2}{\omega_Z^2}\right)\cos(\theta_{d}).
	\end{align}
	
\begin{figure}[t]
		\begin{center}
			\includegraphics[width=3.25 in]{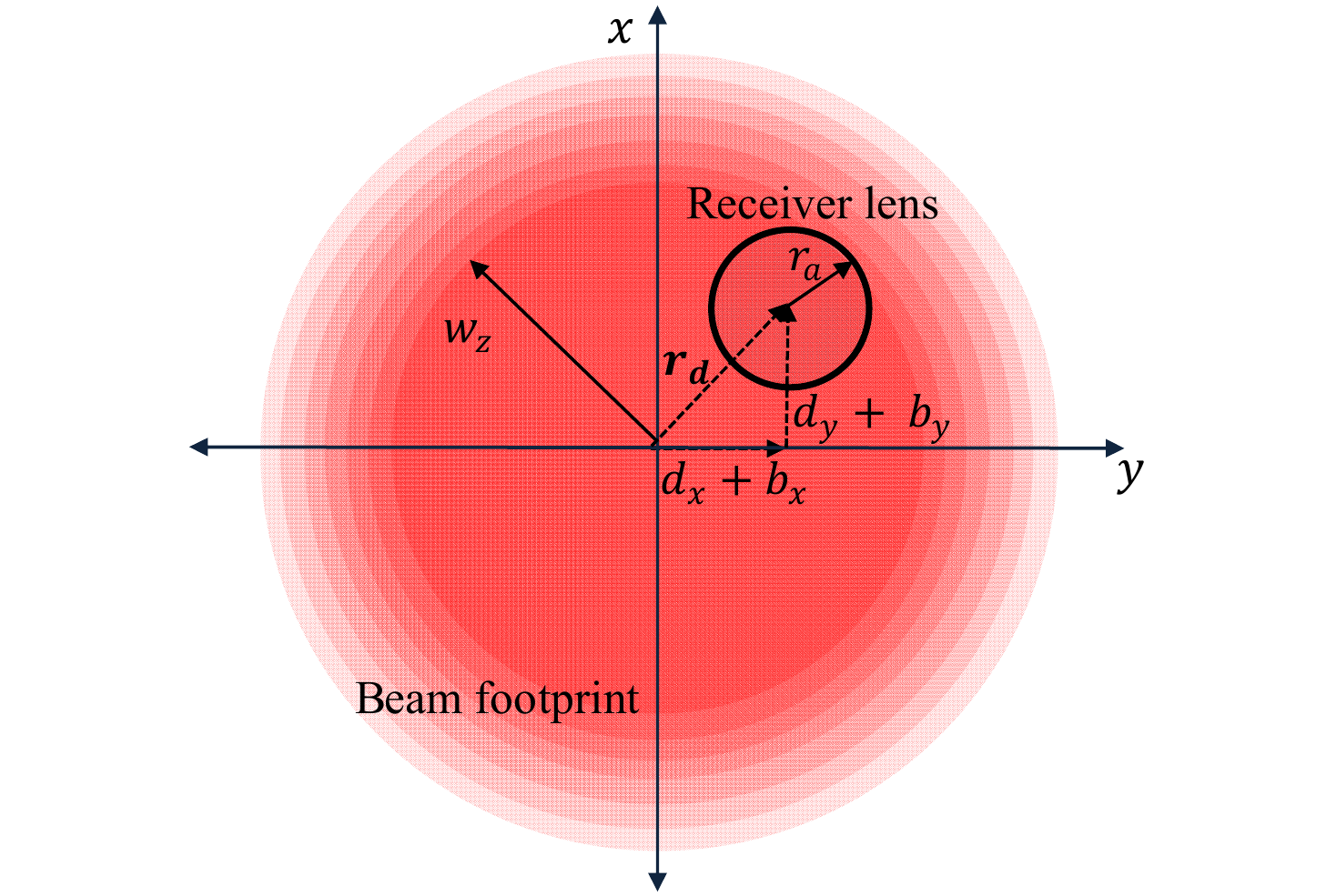}
			\caption{Received optical beam footprint at the receiver aperture for a long-range ground-to-HAP FSO link. Because of the effects of beam wander as well as hovering fluctuations, the center of the beam is randomly deviated from the center of receiver lens.}%Illustration of a typical MR UAV-based FSO communication link.}
			\label{circl}
		\end{center}
		%\vspace*{-6pt}
	\end{figure}

	As shown in Fig. \ref{circl}, by taking beam wander effects and instantaneous position of the HAP into consideration, we have ${||\boldsymbol{r_d}||}=\sqrt{(d_x + b_x)^2+(d_y + b_y)^2} $. Since $d_x$ and $d_y$ are zero-mean Gaussian RVs with variance $\sigma_d^2$, the RV $r_d = ||\boldsymbol{r_d}||$ follows a
	Rayleigh PDF as
	\begin{align}
	\label{zw}
	f_{r_d}(r_d)=\frac{r_d}{\sigma_r^2}\exp\left(-\frac{r_d^2}{2\sigma_r^2}\right),~~~~r_d\geq 0
	\end{align} 
	where $\sigma_r^2 = \sigma_d^2 + \sigma_b^2$. From (\ref{pointing_loss_app}) and (\ref{zw}), and after some manipulations, the PDF of $h_{pl}$ conditioned on $\theta_d$ can be derived as 
	\begin{equation}
	\label{h_poi}
	f_{h_{pl}|\theta_d}(h_{pl}) = C_1^{-C_3}C_3{h_{pl}}^{C_3-1}\cos(\theta_{d})
	\end{equation}	
	where $C_1 = 2\left(\frac{r_a}{\omega_Z^2}\right)^2$, $C_2 = \frac{2}{\omega_Z^2}$, and $C_3 = \frac{1}{2C_2\sigma_r^2}$. 
	%%%%%%%%%%%%%%%%%
	%%%%%%%%%%%%%%%%%
	%\end{IEEEproof}
	%%%%%%%%%%%%%%%%%%%%%%%%%%%%%%%%%
	%%%%%%%%%%%%%%%%%%%%%%%%%%%%%%%%%
	%\begin{figure*}
	%	\normalsize
	%	\begin{align}
	%	\label{fag}
	%	f_{h_{pl}|\theta_d}(h_{pl}) \simeq& \left[1-\exp\left(-\frac{(w_z-r_a)^2}{2\sigma_r^2}\right)\right] 
	%	\times \delta\left(h_{pl}-2\pi\mathcal{C}_1\right)
	%	%%%%%%%%%%%%
	%	+ \exp\left(-\frac{(w_z+r_a)^2}{2\sigma_r^2}\right)  \times \delta\left(h_{pl}\right) \nonumber \\
	%	%%%%%%%%%%%%
	%	&+ \frac{1}{2 \mathcal{C}_2 \sigma_r^2}\exp\left( \frac{ h_{pl}-\mathcal{C}_3}{2 \mathcal{C}_2 \sigma_r^2}  \right)
	%	\times\left[\Pi\left(\frac{h_{pl}}{h_{pg2}}\right) -\Pi\left(\frac{h_{pl}}{h_{pg1}}\right) \right].
	%	\end{align} 
	%	\hrulefill
	%	\vspace*{4pt}
	%\end{figure*}
	%%%%%%%%%%%%%%%%%%%%%%%%%%%%%%%%
	%%%%%%%%%%%%%%%%%%%%%%%%%%%%%%%%

	%Using \cite[eq. (9)]{pointing2014}, 
	Furthermore, the PDF of $h_{ag}=h_{al} h_{at} h_{pl}$ conditioned on $\theta_{d}$ can be obtained as
	\begin{align}
	\label{by}
	f_{h_{ag}|\theta_d}(h_{ag}) =& \int  f_{h_{ag}|\theta_d,h_{at}}(h_{ag})   f_{h_{at}}(h_{at}) dh_{at}  \\
	= & \int  \frac{1}{h_{al} h_{at}}f_{h_{pl}|\theta_d}\left(\dfrac{h_{ag}}{h_{al} h_{at}}\right)   f_{h_{at}}(h_{at}) dh_{at}.\nonumber
	\end{align}
	Substituting \eqref{fg1g}, \eqref{fg1g-LN} and \eqref{h_poi} in \eqref{by} and after some manipulations, we obtain the analytical expressions of $f_{h_{ag}|\theta_d}(h_{ag})$ for low values of $h$ and under both LN and GG atmospheric turbulence models in \eqref{fag4-LN},  and \eqref{fag4}, respectively, where $C_4 = \frac{2C_3(\alpha\beta)^{\frac{\alpha+\beta}{2}}}{h_{al}^{C_3}\Gamma(\alpha)\Gamma(\alpha)}$ and $C_5 =\frac{\alpha+\beta-2C_3-2}{2}$.
	%%%%%%%%%%%%%%%%%%%%%%%%%%%%%%%%%
	%%%%%%%%%%%%%%%%%%%%%%%%%%%%%%%%%
	\begin{figure*}
		\normalsize
		\begin{align}
		\label{fag4-LN}
		f^{LN}_{h_{ag}|\theta_d}(h_{ag}) =&    
		%%%%%%%%%%%%
		\dfrac{C_3{C_1}^{-C_3}h_{ag}^{C_3-1}\cos(\theta_{d})}{2h^{C_3}_{al}\sqrt{2\pi\sigma^2_{Bu}}}
		%%%%%%%%%%%%
		\sqrt{8\pi\sigma^2_{Bu}}\exp\left(8\sigma^2_{Bu}\left(\left(\frac{2C_3+1}{4}\right)^2-\frac{1}{16}\right)\right).
		%%%%%%%%%%%%
		\end{align} 
		\hrulefill
		\vspace*{4pt}
	\end{figure*}
	%%%%%%%%%%%%%%%%%%%%%%%%%%%%%%%%
	%%%%%%%%%%%%%%%%%%%%%%%%%%%%%%%%%
	\begin{figure*}
		\normalsize
		\begin{align}
		\label{fag4}
		f^{GG}_{h_{ag}|\theta_d}(h_{ag}) =&    
		%%%%%%%%%%%%
		\dfrac{2^{2C_5+1}C_1^{-C_3}h_{ag}^{C_3-1}C_4\cos(\theta_d)}{(4\alpha\beta)^{C_5+1}}
		%%%%%%%%%%%%
		\Gamma\left(\frac{2C_5+2+\alpha-\beta}{2}\right)\Gamma\left(\frac{2C_5+2+\beta-\alpha}{2}\right).
		%%%%%%%%%%%%
		\end{align} 
		\hrulefill
		\vspace*{4pt}
	\end{figure*}
	%%%%%%%%%%%%%%%%%%%%%%%%%%%%%%%%
	%%%%%%%%%%%%%%%%%%%%%%%%%%%%%%%%
	%%%%%%%%%%%%%%%%%%%%%%%%%%%%%%%%

	%%%%%%%%%%%%%%%%%%%%%
	%%%%%%%%%%%%%%%%%%%%%
	\begin{figure}[t]
		\centering
		{\includegraphics[width=3.5 in]{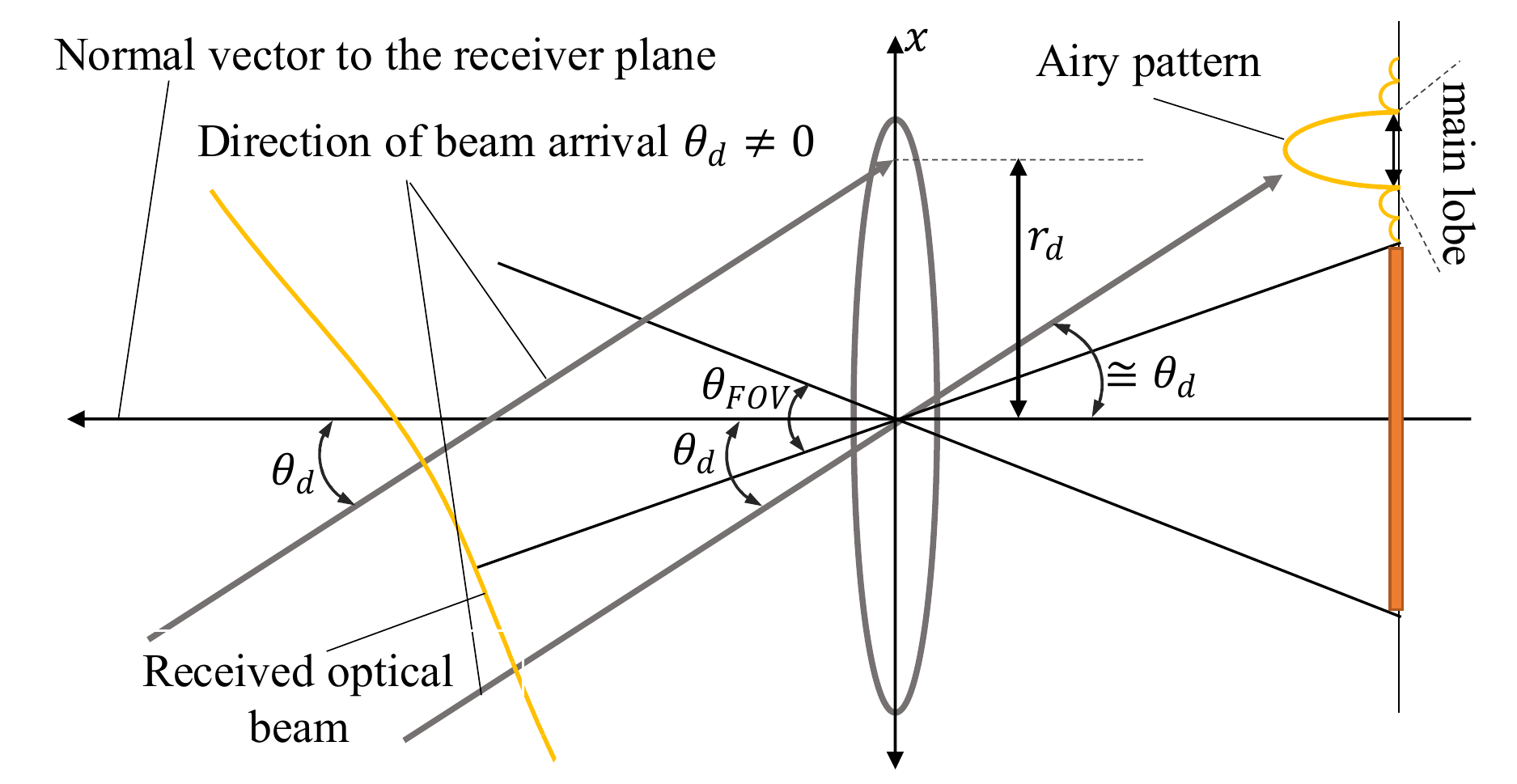}
			\label{4_2}
		}
		\caption{Configurations of received beam and its diffraction pattern in the presence of AOA fluctuations.}
		\label{4}
	\end{figure}
	%%%%%%%%%%%%%%%%%%%%%
	%%%%%%%%%%%%%%%%%%%%%
	%%%%%%%%%%%%%%%%%%%%%%%
	%%%%%%%%%%%%%%%%%%%%%%%
	\subsection{AOA Fluctuations Due to Random Orientation Deviations}
	%%%%%%%%%%%%%%%%%%%%%%%
	%%%%%%%%%%%%%%%%%%%%%%%
	The converging lens focuses collected light to the surface on a circular PD.
	However, due to the random orientation deviations of an aerial node, the angle of arrival of the received beam is deviated from the normal line to the detector area. Indeed, it is crucial to compensate and stabilize the orientation fluctuations of hovering aerial nodes for establishing reliable ground-to-aerial FSO links. More recently, via fast and high accurate stabilization and control
	systems, the degree of angular instability of the hovering
	aerial platforms has been shown on the order of several $ \rm mrad$,  \cite{orsag2017dexterous,lee2017estimation,faessler2017thrust}. 
	Due to this pinpoint accuracy, it is reasonable to assume that the instantaneous misalignment orientations of the aerial platform, i.e., $\theta_{Rx}$ and $\theta_{Ry}$, are sufficiently small and these assumptions  enable us to employ small-angle approximation
	%Thus, assuming that $\theta_{Rx}$ and $\theta_{Ry}$ are sufficiently small, we can use small-angle approximation as ${\rm tan}\left(\theta_{Rx}\right)\simeq \theta_{Rx}$ and ${\rm tan}\left(\theta_{Ry}\right)\simeq \theta_{Ry}$ that let us to approximate $\theta_d^2 = \theta_{Rx}^2 + \theta_{Ry}^2$. According this, distribution of RV $\theta_d$ is well approximated by Rayleigh distribution as $f_{\theta_d}(\theta_d) = $.
	%In this paper, we consider a h
	%The incidence angle relative to the receiver axis is called the AOA of the signal and it is obtained as
	%\begin{align}
	%\label{hjj1}
	%\theta_a = \sqrt{\left(\theta_{tx}+\theta_{rx}\right)^2+\left(\theta_{ty}+\theta_{ry}\right)^2}.
	%\end{align}
	%For an angular deviated receiver as $\theta_d$ respect to the arriving beam direction, the amount of collected power by the receiver lens is
	%Compensating and removing orientation fluctuations of hovering aerial nodes  is one of the most important efforts in the context of the stabilized AP system and the more and more stable HAPs and UAVs have been arising with thanks to the new mechanical and control systems \cite{orsag2017dexterous,lee2017estimation,faessler2017thrust}. 
	%In this paper, we consider highly stable hovering AP nodes which offer high angular stability on the order of  several $ \rm mrad$ thanks to the accuracy of mechanical and control systems \cite{xx}. 
	%Thus, assuming that $\theta_{Rx}$ and $\theta_{Ry}$ are sufficiently small, we can use small-angle approximation 
	as ${\rm tan}\left(\theta_{Rx}\right)\simeq \theta_{Rx}$ and ${\rm tan}\left(\theta_{Ry}\right)\simeq \theta_{Ry}$ \cite{huang2017free}. \textcolor{black}{Hence, the RV $\theta_d = \sqrt{\theta_{Rx}^2 + \theta_{Ry}^2}$ is approximately Rayleigh distributed with PDF \cite{huang2017free}
		\begin{align}
		\label{znw}
		f_{\theta_d}(\theta_d)=\frac{\theta_d}{\sigma_o^2}\exp\left(-\frac{\theta_d^2}{2\sigma_o^2}\right),~~~~\theta_d\geq 0.
		\end{align} 
		%As mentioned in Section II-D, some previous works only take into account the main lobe of Airy pattern to determine signal detection at a limited-FOV FSO receiver as described by (27). Under such simplified assumption, an analytical expression for outage probability can be easily achieved by substituting (27) into (38) [21].
		%
		%the power contained in the side lobes of Airy pattern is non-negligible which means that we can still collect a significant amount of power even though the AOA is outside the receiver FOV.
		%
		%Since the fraction of power in side lobes is much smaller than that in main lobe,
		%
		%
		% the main lobe of Airy pattern to determine signal detection %%%% Under such simplified assumption,
		%In the previous works \cite{dabiri2018channel,gagliardi1995optical,arnon2003effects}, it is assumed that the fraction of power in side lobes of Airy pattern is much smaller than that in main lobe. Therefore, one can argue that the effect of power in side lobes is insignificant. 
		When an optical beam having small angle of deviation  $\theta_d$ from the normal vector of aperture plane is passed through a lens, the outside angle of beam will be approximately unchanged \cite{gagliardi1995optical}. Therefore, as depicted in Fig. \ref{4}, AOA fluctuations cause shifted
		diffracted patterns which can attenuate the amount of received optical power at the PD.}
	
	\textcolor{black}{
		For a circular shape aperture, Airy pattern at the PD consists of a bright disc at the center of the pattern (main-lobe) surrounded by concentric bright and dark rings (side-lobes) \cite{born2013principles}. Let $h_{af}$ denote the
		fraction of power collected by the detector to the power
		incident in the aperture \cite{huang2017free}. 
		% at high transmit power regime, however, the receiver can still collect the power of the side-lobes  despite the fact that
		%the main-lobe lies outside the FOV of the receiver due to AOA fluctuations. \cite{born2013principles}.  
		Therefore, to calculate the amount of $h_{af}$, it is essential to explore what fraction of total incident power is contained within the main-lobe and side lobes of the Airy pattern. Let $L(\psi)$ be the fraction of total power of the Airy pattern contained in a circle with radius $\psi$ to the power incident in the aperture. From \cite[8.5 (18)]{born2013principles}, $L(\psi)$ is obtained as
		\begin{align}
		\label{fraction_power}
		L(\psi) = 1 - J_0^2\bigg(\frac{\pi\psi}{\lambda}\bigg) - J_1^2\bigg(\frac{\pi\psi}{\lambda}\bigg)
		\end{align}
		where $J_n(\cdot)$ denotes Bessel function of the first kind with order $n$.
		\begin{figure}[t]
			\begin{center}
				\includegraphics[width=3.5 in]{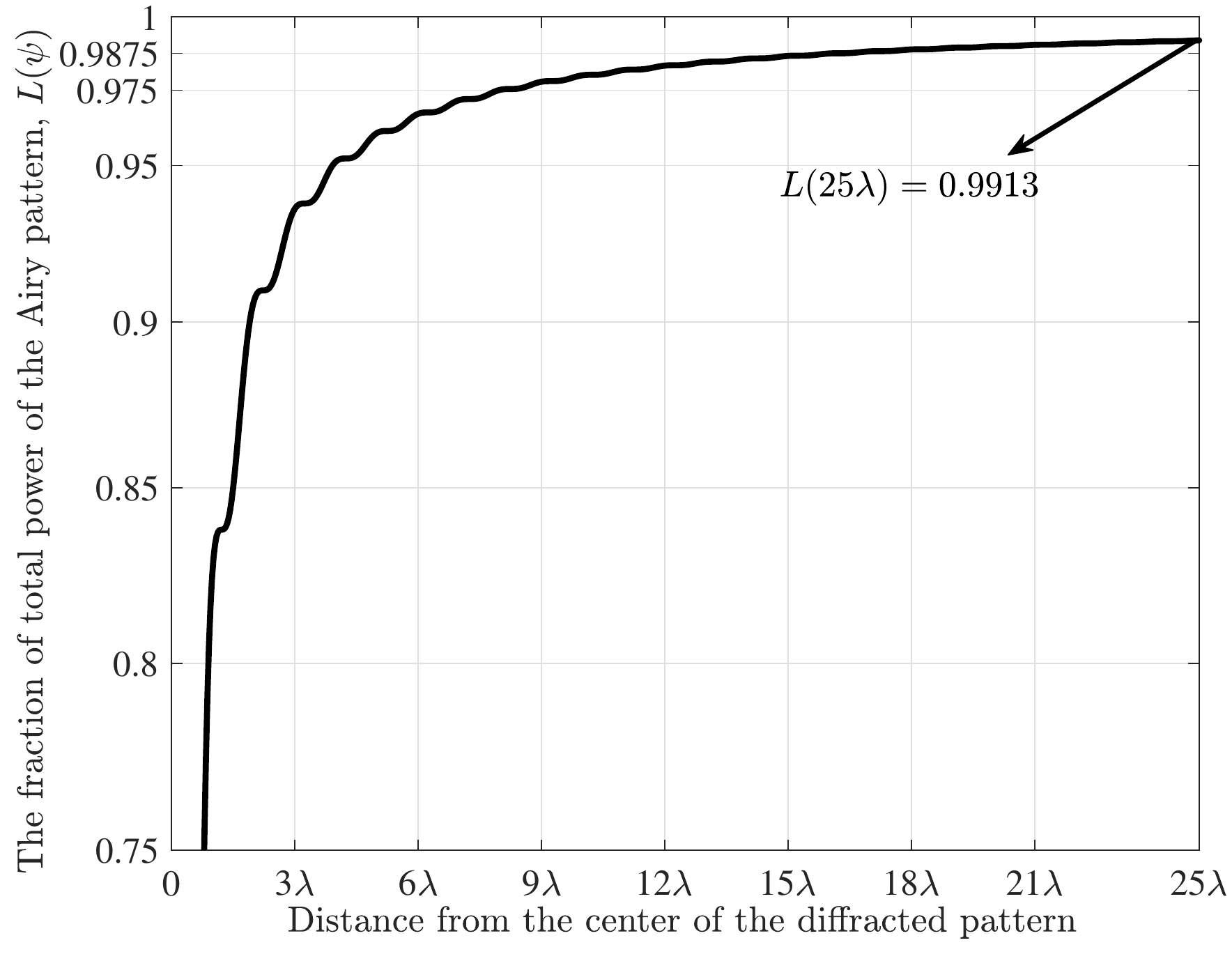}
				\caption{\textcolor{black}{The fraction of the power of Airy pattern at the PD contained in a circle with radius $\psi$.}}%Illustration of a typical MR UAV-based FSO communication link.}
				\label{fraction_power_Airy_pattern}
			\end{center}
			%\vspace*{-6pt}
		\end{figure}
		To obtain detailed insight into the distribution of the power of Airy pattern at the PD, we plot $L(\psi)$ versus $\psi$ in Fig. \ref{fraction_power_Airy_pattern}. As we can observe from this figure, more than 99 percent of the power of the Airy pattern is contained within a circle with radius $25\lambda$,  and it is much smaller than the conventional sizes of a PD, which is typically on order of $\rm mm$ \cite{gagliardi1995optical}. Therefore, when
		the received laser beam lies inside the receiver FoV, i.e., when $\theta_d \leq \theta_{FOV}$, the total power of the incident beam is captured by the PD. Meanwhile, as thoroughly discussed in \cite{huang2017free}, ignoring the effect of side-lobes may lead to an incorrect outage probability floor  when the receiver FOV is not large enough and the AOA lies outside of the FOV. Accordingly, an accurate approximate expression was proposed for the fading introduced by AOA fluctuations $h_{af}$  for a receiver with a circular PD with radius $r_p$  as \cite{huang2017free}
		\begin{align}
		\label{haf}
		h_{af} = 
		\begin{cases}
		L(r_p) \,\,\,\,\,\,\,\,\,\,\,\,\,\,\,\,\,\,\,\,\,\,\,\,\,\,\,\,\,\,\,\,\,\,\,\,\,\,\,\,\,\,\,\,\,\,\,\,\,\,\,\,\,\,\,\,\,\,\,\,\,\,\,\,\,\,\,\,\,\,\,\, \theta_d \leq \theta_\textrm{FOV}, \\
		\frac{r_p}{4r_d}\bigg(L(r_d + r_p) - L(r_d - r_p)\bigg) \,\,\,\,\,\,\,\,\,\, \text{otherwise}. 
		\end{cases}
		\end{align}
		From \eqref{haf} and \eqref{xp}, the PDF of the considered ground-to-HAP link is obtained as
		\begin{align}
		\label{xp11}
		f_h(h) =\! \int_0^{\infty}\!\!\!\! h_{af}f_{h_{ag}|\theta_d}(h) f_{\theta_d}(\theta_d) d\theta_d.
		\end{align}
		Nevertheless, the system model of \cite{huang2017free} concerns terrestrial FSO links with a narrow receiver FOV, i.e., $0.1 \sim 0.2 \textrm{~mrad}$. However, as we will show in the numerical result section, due to the large effect of hovering fluctuations of the HAPs, the optimal FOV for such receivers is on the order of several tens of milli-radians. {To make the analysis tractable, we make an approximation to the values of $h_{af}$ by ignoring the effect of sidelobes as follows
			\begin{align}
			\label{xp}
			\tilde{h}_{af}\simeq \Pi\left( \frac{\theta_d}{\theta_{FOV}}\right)
			\end{align}
			where the gate function $\Pi(\cdot)$ is  defined as
			\begin{equation}
			\label{vcv1}
			\Pi(x) = \left\{
			\begin{array}{rl}
			1 &~ \text{if }~ x < 1\\
			0 &~ \text{if }~ x > 1.
			\end{array} \right.
			\end{equation}
			As it will be shown in the numerical results, the effect of sidelobes is small in such cases as no error floors are observed within the the range of interest of the system performance. Therefore, eq. \eqref{xp} is a reasonable approximation for systems with large FOV which is the case for HAPs.}
	}Thus, eq. \eqref{xp11} can be rewritten as
	\begin{align}
	\label{xp1}
	f_h(h) =\! \int_0^{\theta_{FOV}}\!\!\!\! f_{h_{ag}|\theta_d}(h) f_{\theta_d}(\theta_d) d\theta_d 
	\! +\! \delta(h) \! \int_{\theta_{FOV}}^\infty\!\! f_{\theta_d}(\theta_d) d\theta_d.
	\end{align}
	Nevertheless, it can be cumbersome to evaluate the integral equation in \eqref{xp1}. To have a more tractable analytical channel model, for small values of $\theta_d$, we can use small angle approximation as $ \cos(\theta_d) \simeq 1$. Accordingly, for small values of $h$, eq. \eqref{xp1} can be simplified as \eqref{fh-LN} under LN atmospheric turbulence, and as \eqref{xp2} under GG atmospheric turbulence. Furthermore, based on \eqref{fh-LN} and \eqref{xp2}, closed-form expressions for the outage probabilities  are derived in \eqref{outage-LN} and \eqref{outage_probability}, respectively, for LN and GG atmospheric turbulence models.
	%$\mathcal{C}'_3 = \mathcal{C}'_2 \left(\pi w_z r_a - r_a^2+w_z^2 \right)$,
	%%%%%%%%
	%$\mathcal{C}'_2 = \left( \frac{\mathcal{C}'_1}{w_z r_a} \right)$,
	%%%%%%%%
	%$\mathcal{C}'_1=\frac{ r_a^2     }     {4\pi\left( 1-\cos\left(\frac{\theta_\textrm{div}}{2}\right)\right)Z^2}$, 
	%$h'_{pg1}=\mathcal{C}'_1  \left(\pi - \frac{2r_a^2+2r_a w_z}{w_z r_a} \right)$ and
	%$h'_{pg2}=\mathcal{C}'_1    \left(\pi - \frac{2r_a^2-2r_a w_z}{w_z r_a} \right)$. 
	%%%%%%%%%%%%%%%%%%%%%%%%%%%%%%%%%
	%%%%%%%%%%%%%%%%%%%%%%%%%%%%%%%%%
	\begin{figure*}
		\normalsize
		\begin{align}
		\label{fh-LN}
		f^{LN}_{h}(h) &\simeq 
		%%%%%%%%%%%%
		%%%%%%%%%%%%
		\underbrace{		\dfrac{C_3{C_1}^{-C_3}h^{C_3-1}}{2h^{C_3}_{al}\sqrt{2\pi\sigma^2_{Bu}}}
			\sqrt{8\pi\sigma^2_{Bu}}\exp\left(8\sigma^2_{Bu}\left(\left(\frac{2C_3+1}{4}\right)^2-\frac{1}{16}\right)\right)\left(1-e^{-\frac{\theta_{FOV}^2}{2\sigma_o^2}}\right)}_{f'_h(h)}  \\
		& +\underbrace{    e^{-\frac{\theta_{FOV}^2}{2\sigma_o^2}} \delta(h) }_{f^0_h(h)}. \nonumber
		%%%%%%%%%%%%
		\end{align} 
		\hrulefill
		\vspace*{4pt}
	\end{figure*}
	%%%%%%%%%%%%%%%%%%%%%%%%%%%%%%%%
	%%%%%%%%%%%%%%%%%%%%%%%%%%%%%%%%%
	\begin{figure*}
		\normalsize
		\begin{align}
		\label{xp2}
		f^{GG}_{h}(h) &\simeq 
		%%%%%%%%%%%%
		%%%%%%%%%%%%
		\underbrace{	\dfrac{2^{2C_5+1}C_1^{-C_3}h^{C_3-1}C_4}{(4\alpha\beta)^{C_5+1}}		 	\Gamma\left(\frac{2C_5+2+\alpha-\beta}{2}\right)\Gamma\left(\frac{2C_5+2+\beta-\alpha}{2}\right)\left(1-e^{-\frac{\theta_{FOV}^2}{2\sigma_o^2}}\right)}_{f'_h(h)}  \\
		& + \underbrace{    e^{-\frac{\theta_{FOV}^2}{2\sigma_o^2}} \delta(h) }_{f^0_h(h)}. \nonumber
		%%%%%%%%%%%%
		\end{align} 
		\hrulefill
		\vspace*{4pt}
	\end{figure*}
	%%%%%%%%%%%%%%%%%%%%%%%%%%%%%%%%
	%%%%%%%%%%%%%%%%%%%%%%%%%%%%%%%%%
	%%%%%%%%%%%%%%%%%%%%%%%%%%%%%%%%%
	\begin{figure*}
		\normalsize
		\begin{align}
		\label{outage-LN}
		\mathbb{P}^{LN}_{\textrm{out}} &=   {    e^{-\frac{\theta_{FOV}^2}{2\sigma_o^2}}}
		%%%%%%%%%%%%
		+  
		%%%%%%%%%%%%
		{		\dfrac{C_3{C_1}^{-C_3}h_{\textrm{th}}^{C_3}}{2C_3h^{C_3}_{al}\sqrt{2\pi\sigma^2_{Bu}}}
			\sqrt{8\pi\sigma^2_{Bu}}\exp\left(8\sigma^2_{Bu}\left(\left(\frac{2C_3+1}{4}\right)^2-\frac{1}{16}\right)\right)\left(1-e^{-\frac{\theta_{FOV}^2}{2\sigma_o^2}}\right)}. \nonumber \\
		%%%%%%%%%%%%
		\end{align} 
		\hrulefill
		\vspace*{4pt}
	\end{figure*}
	%%%%%%%%%%%%%%%%%%%%%%%%%%%%%%%%
	%%%%%%%%%%%%%%%%%%%%%%%%%%%%%%%%%
	\begin{figure*}
		\normalsize
		\begin{align}
		\label{outage_probability}
		\mathbb{P}^{GG}_{\textrm{out}} &=    
		%%%%%%%%%%%%
		e^{-\frac{\theta_{FOV}^2}{2\sigma_o^2}} + 
		%%%%%%%%%%%%
		\dfrac{2^{2C_5+1}C_1^{-C_3}h_\textrm{th}^{C_3}C_4}{(4\alpha\beta)^{C_5+1}C_3}		 	\Gamma\left(\frac{2C_5+2+\alpha-\beta}{2}\right)\Gamma\left(\frac{2C_5+2+\beta-\alpha}{2}\right)\left(1-e^{-\frac{\theta_{FOV}^2}{2\sigma_o^2}}\right).
		%%%%%%%%%%%%
		\end{align} 
		\hrulefill
		\vspace*{4pt}
	\end{figure*}
	%%%%%%%%%%%%%%%%%%%%%%%%%%%%%%%%
	%%%%%%%%%%%%%%%%%%%%%%%%%%%%%%%%
	%From \eqref{xp2}, $f_h(h)$ consists of two terms, i.e., $f_h(h)=f_h(h=0)+f_h(h>0)$ where $f_h(h=0)=f^0_h(h)$ and $f_h(h>0)=f'_h(h)+f''_h(h)$.
	%Moreover, for large values of $w_z$, the term $f''_h(h)$ is negligible compared to $f'_h(h)$, thus \eqref{xp2} is simplified to 
	%\begin{align}
	%\label{xp3}
	%	f_{h}(h) \simeq f^0_h(h) +  f'_h(h).
	%\end{align}
	%%%%%%%%%%%%%%%%%%%%%
	%%%%%%%%%%%%%%%%%%%%%
	\begin{figure}
		\centering
		{\includegraphics[width=3.5 in]{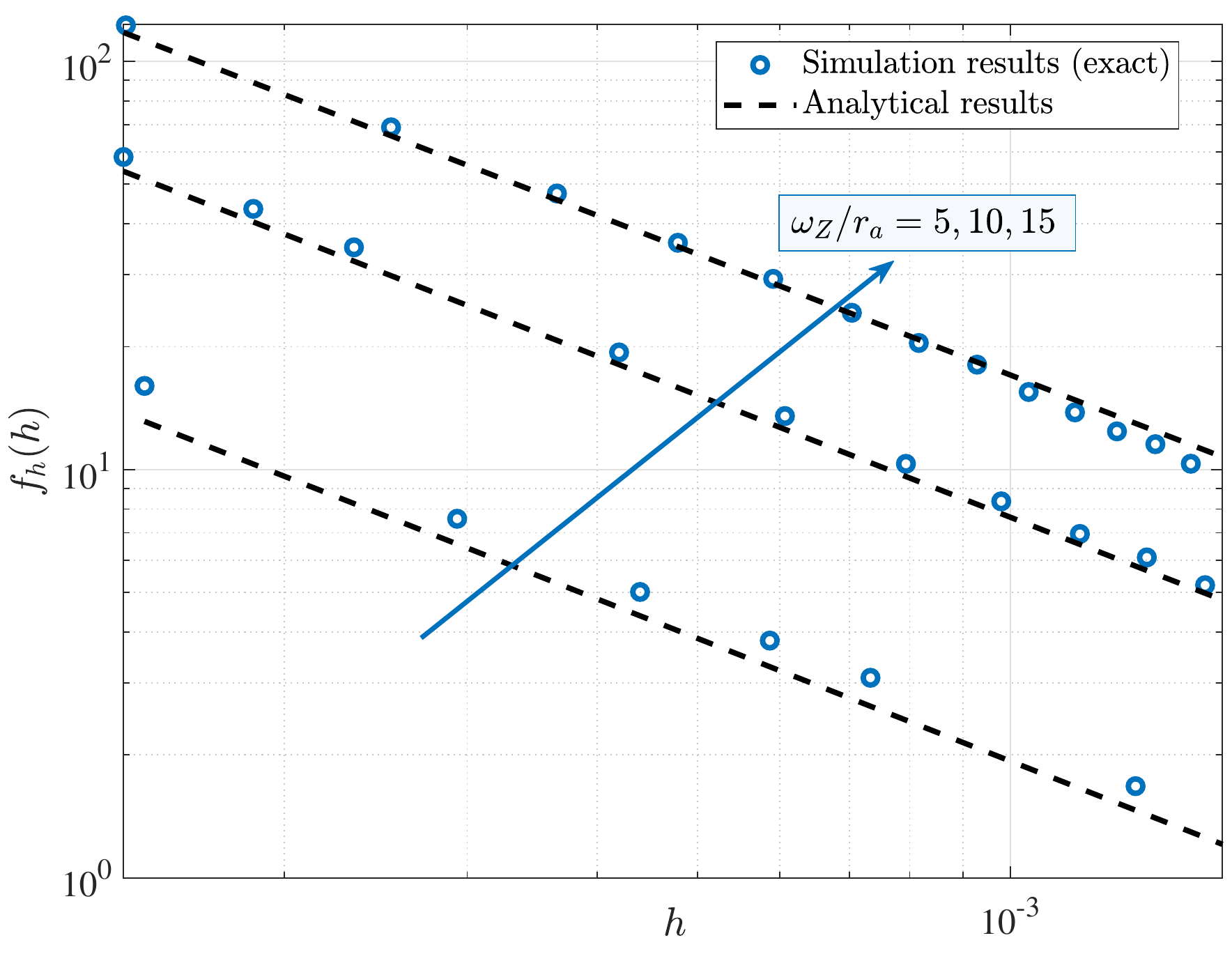}
			\label{PDF_1}
		}
		\caption{Channel distribution $f_h(h>0)$, for different values of  $w_Z/r_a$,  under Log-normal atmospheric turbulence.}
		\label{df}
	\end{figure}
	%%%%%%%%%%%%%%%%%%%%%
	%%%%%%%%%%%%%%%%%%%%%
	%%%%%%%%%%%%%%%%%%%%%
	%%%%%%%%%%%%%%%%%%%%%
	\begin{figure}
		\centering
		{\includegraphics[width=3.5 in]{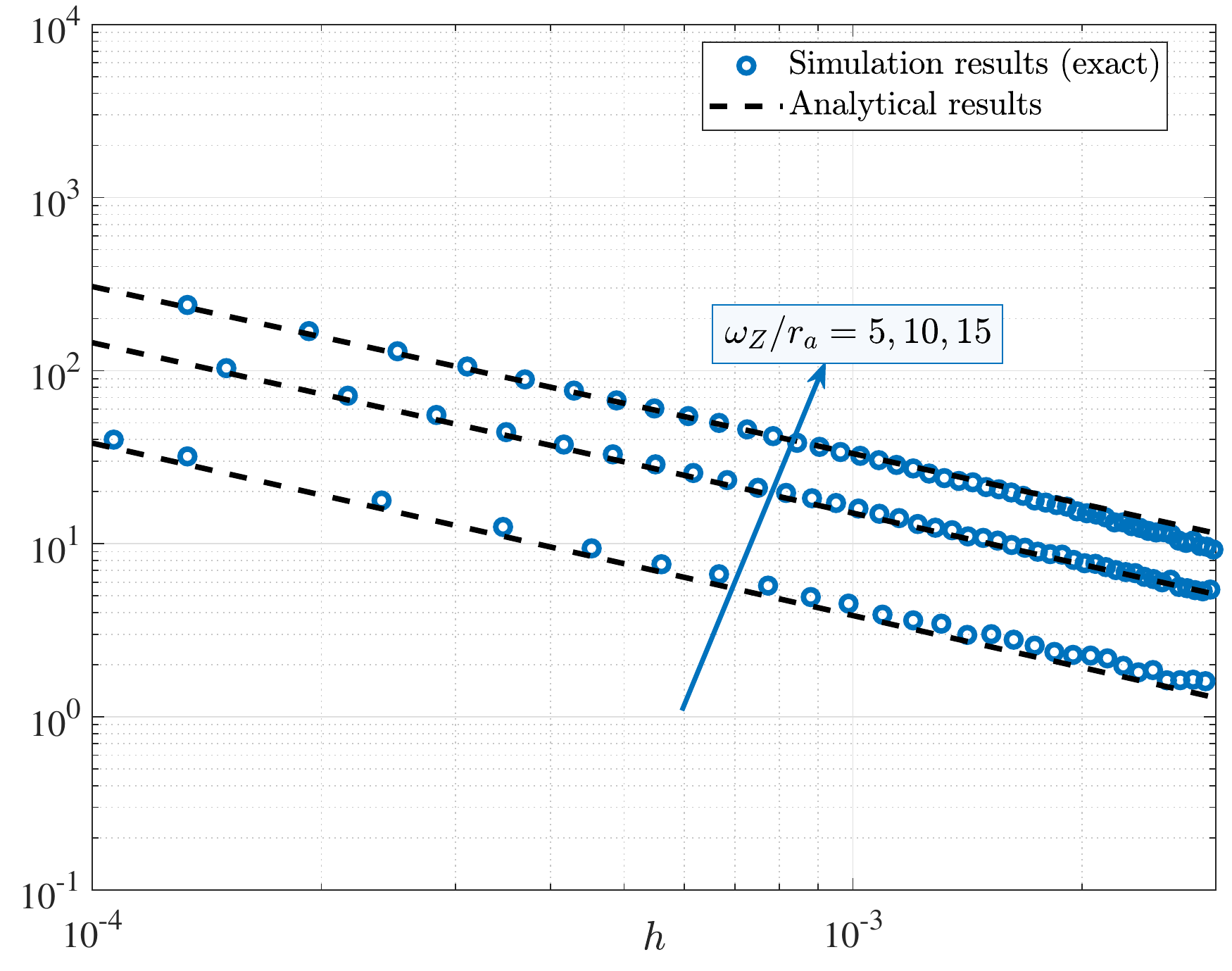}
			\label{PDF_2}
		}
		\caption{Channel distribution $f_h(h>0)$, for different values of  $w_Z/r_a$,  under Gamma-Gamma atmospheric turbulence.}
		\label{df1}
	\end{figure}
	\section{Simulation Results and Analysis}
	%%%%%%%%%%%%%%%%%%%%%%%%%%%%%%%%%%%%%%
	%%%%%%%%%%%%%%%%%%%%%%%%%%%%%%%%%%%%%%
	In this section, we illustrate the analytical results derived
	in the previous sections. We evaluate the link performance in terms of the outage probability and study the impact of different parameters, i.e., the FoV of the receiver and its orientation deviations, and the beam spot size at the transmitter, on the performance of the ground-to-HAP link. Meanwhile,  the accuracy of the derived analytical expressions is corroborated by Monte Carlo simulations using over $4\times10^6$ independent runs.
	%channel model expression for different values of the parameters related to aperture area and beam-width at the receiver.  
	%Moreover, we continue to evaluate the link performance in terms of the outage probability and study the impact of different parameters, i.e., the FoV of the receiver and its orientation deviations, as well as the beam spot size at the transmitter, on the performance of the ground-to-HAP link.
	Simulations are performed based on the practical values of
	the parameters outlined in Table \ref{parameters} \cite{ghassemlooy2012optical}. Since the height of the receiver, $H$, is much larger than the height of the optical transmitter, $h_0$, without loss of generality, we neglect the height of the transmitter for calculating the link length $Z$ (i.e., we assume that $h_0$ is equal to zero). \textcolor{black}{We consider a strong turbulence model for atmospheric turbulence, i.e., GG model; however, the results and discussions can be readily developed to weak turbulence scenarios by applying the LN model and its related expressions  in Section III.}
	
	\begin{table}
		\caption{Nominal Values Used for the Numerical Results } % title of Table
		\centering % used for centering table
		%\resizebox{.7\hsize}{!}{
		\begin{tabular}{l c r} % centered columns (3 columns)
			\hline\hline \\[-1.2ex]%inserts double horizontal lines\\
			Name & Parameter & Value \\ [.5ex] % inserts table
			%heading
			\hline\hline \\[-1.2ex]% inserts single horizontal line
			Optical wavelength                        & $ \lambda $           &$ 1550 $ nm \\[1ex] 
			PD responsibility              & $ \eta $         & $ 0.9 $ \\[1ex]
			%Plank\textquotesingle s Constant&$ h_{p} $        &$ 6.6 \times 10^{-34} $  \\[1ex]
			Optical bandwidth of the receiver                    &$ B_o $      &$ 10 $ nm  \\[1ex]               
			%	Optical Frequency           & $ \nu $         & $ 1.93 \times 10^{14} $ \\[1ex]              
			Receiver electrical bandwidth&$ B_e $    & $ 1 $ GHz \\[1ex]                
			Spectral radiance of \\the background radiation at $\lambda$      & $ N_b(\lambda) $       & $10^{-3}$ W/$\text{cm}^2$-m-srad \\[1ex]                
			HAP zenith angle  &$ \zeta $        & $40^\circ$   \\[1ex]                
			Link length             &$ Z $        & $ 20 $ km \\[1ex]  
			The RMS of wind speed               &$ V_{\omega} $            & $ 21  $ m/s \\[1ex]               
			Electron charge         & $ e $           & $ 1.6 \times 10^{-19} $ \\[1ex]               
			Refractive index structure \\at the ground      & $ C_n^2(0) $       & $1.7 \times 10^{-13}$ $\text{m}^{-2/3}$ \\[1ex]
			\hline\hline %inserts single line
		\end{tabular}
		%}
		\label{parameters} % is used to refer this table in the text
	\end{table}
	%we consider typical system parameters as follows. The optical wavelength $\lambda = 1550$ nm, photo-detector responsibility $\eta$ = 0.9, the optical bandwidth of the receiver $B_o$ = 10 nm, electrical bandwidth $B_e$ = 1 GHz, spectral radiance $N_b(\lambda)$ = $10^{-3}$ W/$\text{cm}^2$-m-srad, the HAP zenith angle $\zeta = 40^\circ$, the link length $Z$ = 20 km, the rms of wind speed $V_w$ = 21 m/s, and the nominal value of $C_n^2(0)$, i.e., $S_t =$ $1.7 \times 10^{-13}$ $\text{m}^{-2/3}$ \cite{ghassemlooy2012optical}. 

	First, we show in Figs. \ref{df} and \ref{df1}, respectively, the channel distribution under both LN and GG atmospheric turbulence for different values of $w_Z/r_a$. From these two figures we observe that the accuracy of the derived analytical channel model depends on the ratio of $w_Z$ and $r_a$. Since aperture radius in FSO systems is on the order of several centimeters and also, for long-range ground-to-HAP links, $w_Z$ is on the order of several meters, the proposed analytical expression for the channel model of such links achieves acceptable level of accuracy. Therefore, the system
	performance metrics for a ground-to-HAP FSO link, e.g., outage probability, and bit error rate can be analytically developed without resorting to time-consuming simulations. 
	
\begin{figure}[t]
		\begin{center}
			\includegraphics[width=3.50 in]{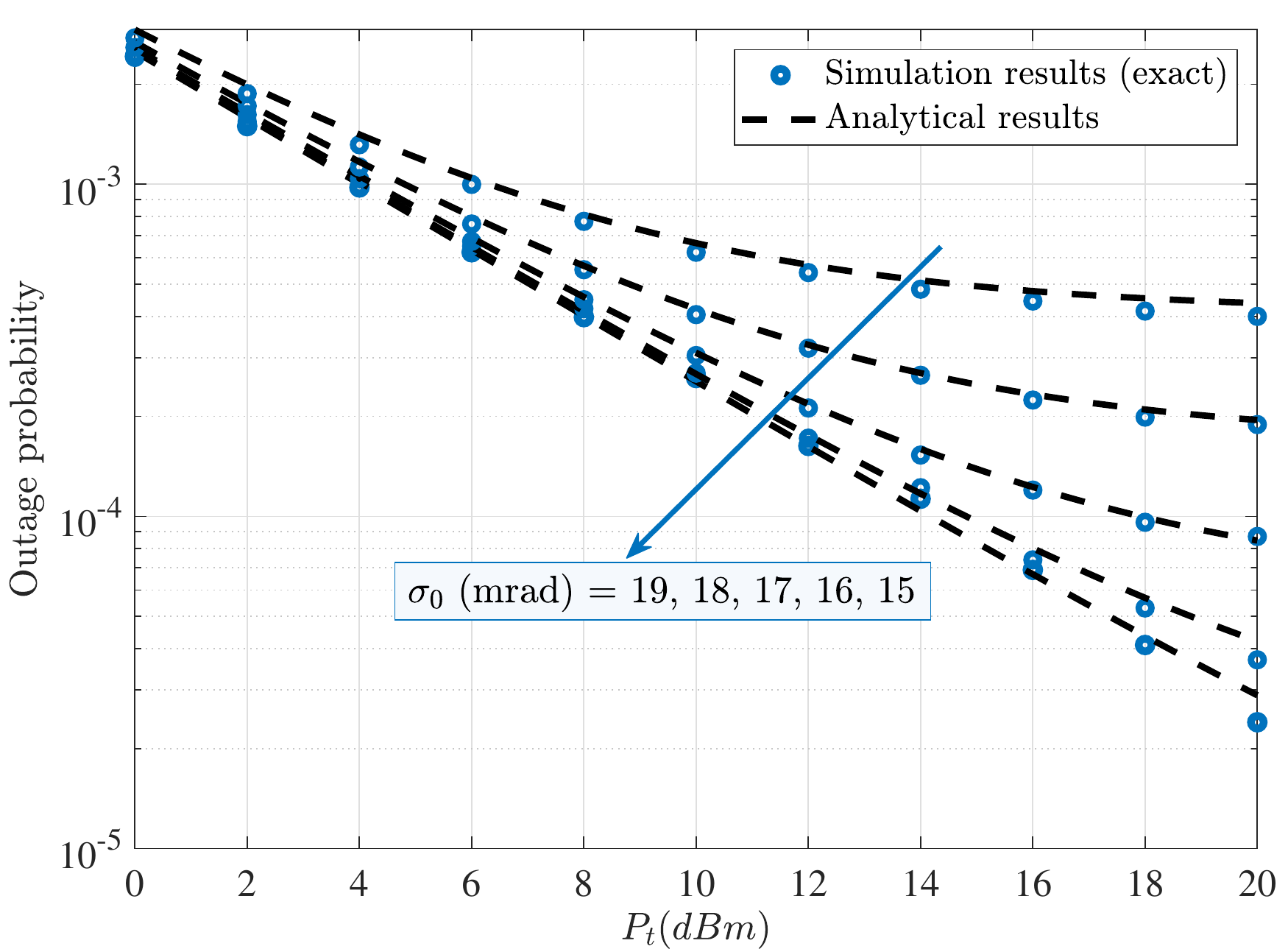}
			\caption{Outage probability versus $P_t$ for $w_Z/r_a = 20$, $\theta_\textrm{FOV} = 75 \textrm{~mrad}$, $\sigma_d = 0.4 \textrm{~m}$, and different values of the orientation deviations $\sigma_0$.}%Illustration of a typical MR UAV-based FSO communication link.}
			\label{outage_vs_orinetation}
		\end{center}
		%\vspace*{-6pt}
	\end{figure}

\begin{figure}[t]
		\begin{center}
			\includegraphics[width=3.5 in]{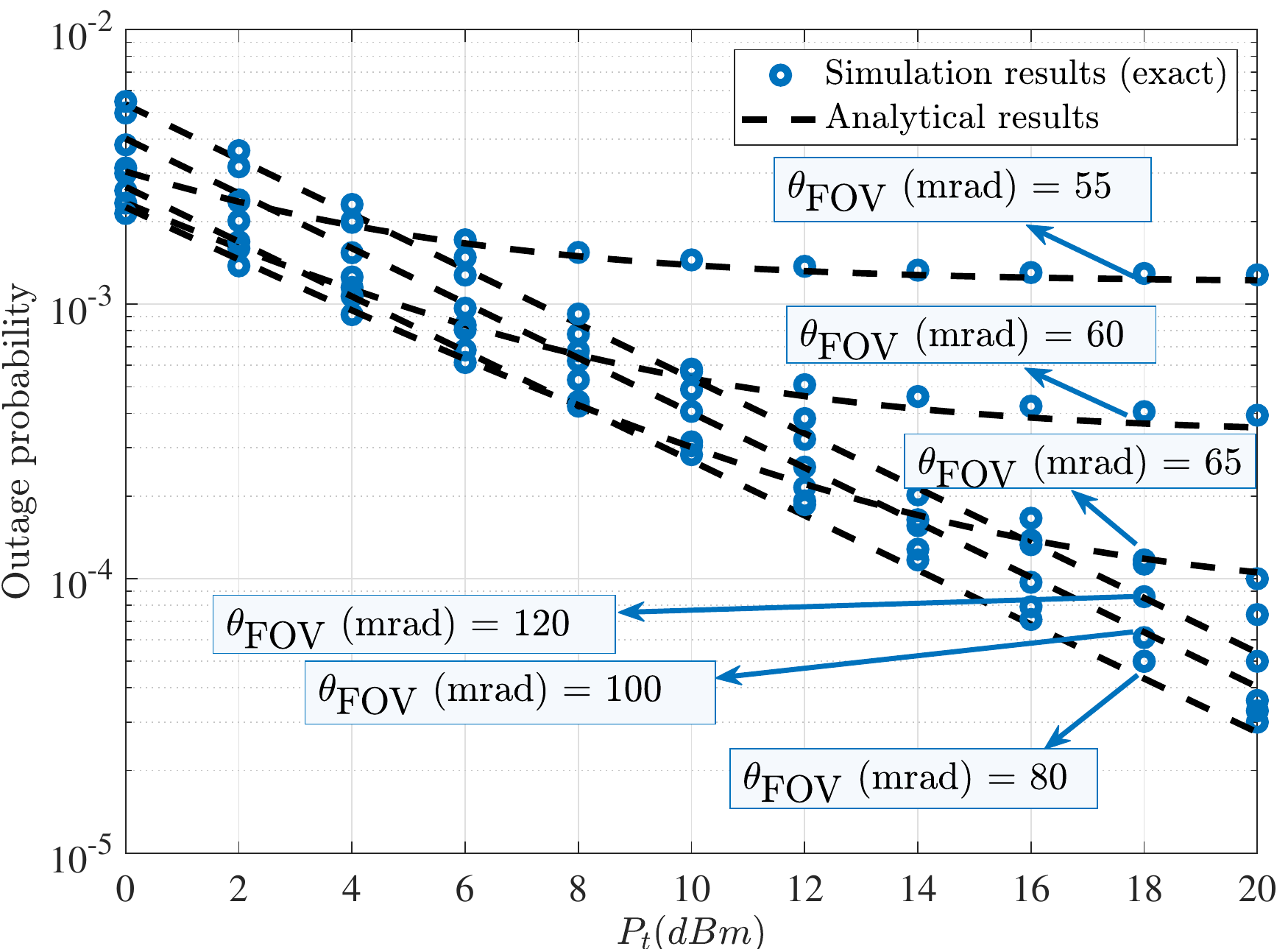}
			\caption{Outage probability versus $P_t$ for $w_Z/r_a = 20$, $\sigma_0 = 15 \textrm{~mrad}$, $\sigma_d = 0.4 \textrm{~m}$, and different values of the receiver FOV  $\theta_\textrm{FOV}$.}%Illustration of a typical MR UAV-based FSO communication link.}
			\label{outage_FoV}
		\end{center}
		%\vspace*{-6pt}
	\end{figure}

\begin{figure}[t]
		\begin{center}
			\includegraphics[width=3.5in]{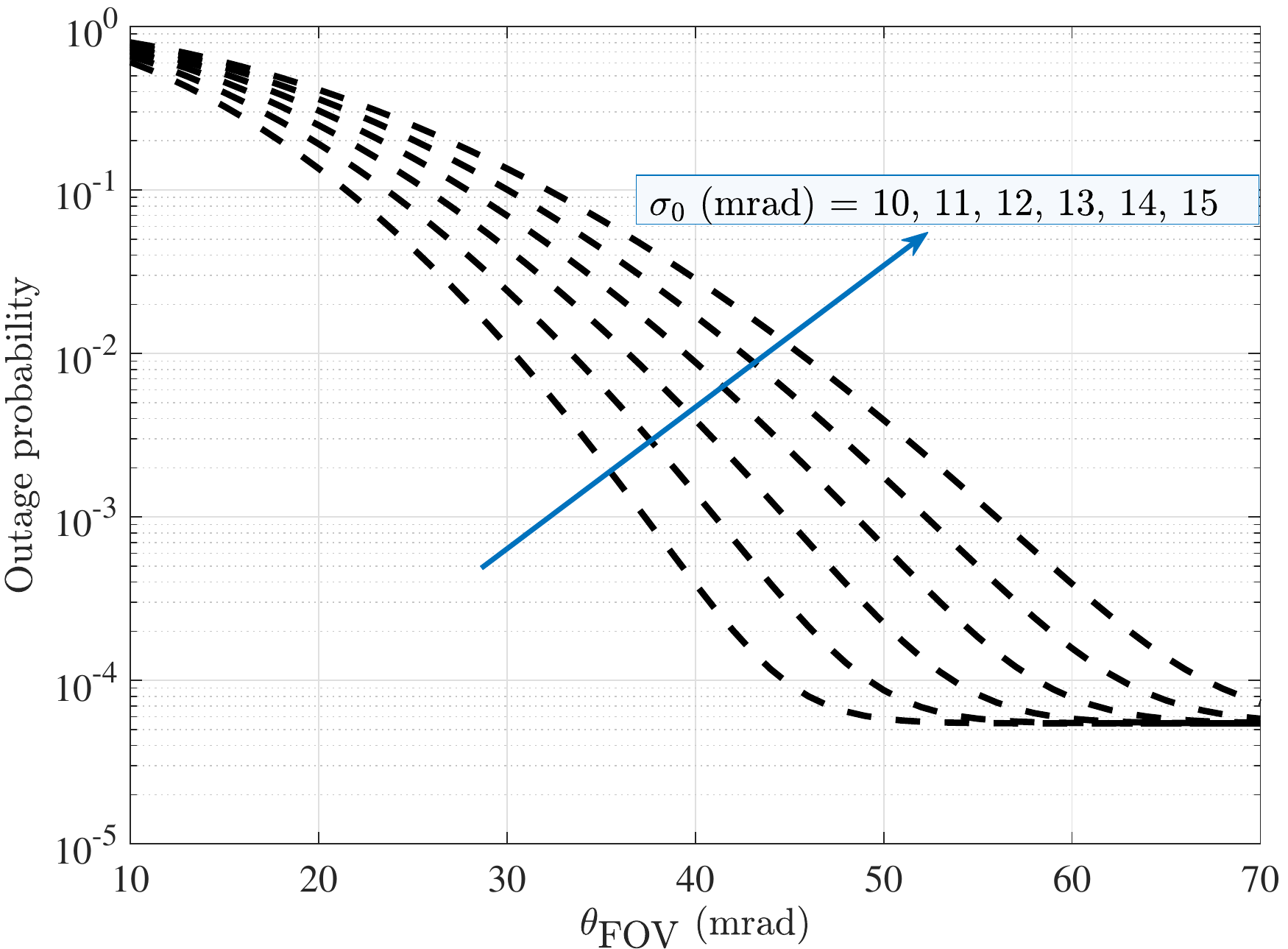}
			\caption{Outage probability versus $\theta_{\textrm{FOV}}$ for $W_Z/r_a = 20$,  $\sigma_d = 0.4 \textrm{~m}$, and different values of the orientation deviations $\sigma_0$.}
			\label{outage3da}
		\end{center}
		%\vspace*{-6pt}
	\end{figure}

	To broaden our understanding about the impact of orientation deviations due to hovering fluctuations of the receiver on
	the link performance, we plot outage probability versus $P_t$ for different values of $\sigma_0^2$  in Fig. \ref{outage_vs_orinetation}.  As shown, an exact match between the analytical and simulation-based results can
	be observed, which validates the accuracy of the derived analytical expression for the outage probability. Also, as we can observe from this figure, the performance of such links  largely depends
	on the AOA fluctuations due to random orientation deviations of the HAP, and, as expected, the outage probability increases when AOA fluctuations at the receiver side is increased.
	%%%%%%%%%%%%%%%%
	%%%%%%%%%%%%%%%%%
	However, such performance degradation can be improved by increasing the receiver FOV. On the other hand, an increase of  FOV of the receiver also increases the amount of undesired background noise, which adversely affects the link performance. To study the inherent tradeoff in optimizing the receiver FOV, we plot  outage probability 
	versus $P_t$ for different values of the receiver FOV and for $\sigma_0^2 = 15 \textrm{~mrad}$ in Fig. \ref{outage_FoV}. As shown, from Fig. \ref{outage_FoV}, for  given values of AOA fluctuations, the link performance is  sensitive to the amount of the receiver FOV. Moreover, from Fig. \ref{outage3da}, one can realize that increasing the amount of FOV does not necessarily improve the system performance. Figs. \ref{outage_FoV}  and \ref{outage3da}  demonstrate the importance of designing optimal receiver FOV to alleviate the impacts of AOA fluctuations on the performance of the ground-to-HAP FSO links.
	%%%%%%%%%%%%%%%%
	%%%%%%%%%%%%%%%%%
	%%%%%%%%%%%%%%%%
	%%%%%%%%%%%%%%%%%
	The optimal $\theta_\textrm{FOV}$ values to achieve minimum outage probability over the considered ground-to-HAP FSO link for different values of $\sigma_0$ are provided in Table \ref{FOV-OPT}. The analytical results of this table are obtained by differentiating \eqref{outage_probability} with respect to $\theta_\textrm{FOV}$
	and setting the result equal to zero. Meanwhile,  Table \ref{FOV-OPT} confirms the accuracy of the proposed analytical
	expressions for outage probability, making it easy to study and design such HAP-based FSO communication links under the degree of instability of the HAP (i.e., the amount of hovering fluctuations).
	%%%%%%%%%%%%%%%%%%%%%%%%%%%%%%%%%%%%%%%
	\begin{table}
		\def\tablename{Table}
		\centering
		\caption{Optimal values of $\theta_{FOV}$ to achieve minimum outage probability over ground-to-HAP link for different values of $\sigma_0$.}
		%\resizebox{.85\hsize}{!}{%
		%%%%%%%%%%%%%%%%%%%%%%%%%%%%%%%%%%%%
		%%%%%%%%%%%%%%%%%%%%%%%%%%%%%%%%%%%
		\begin{tabular}{|c|c|c|c|c|}
			\hline
			& \multicolumn{4}{c|}{$P_t = 5$ dBm, $w_Z/r_a =$ 10, and $\sigma_d$ = 0.2 m}                                          \\ \hline
			\multirow{2}{*}{$\sigma_{0}$ (mrad)} & \multicolumn{2}{c|}{\multirow{2}{*}{ $\theta_{FOV}$ (mrad)} } & \multicolumn{2}{c|}{Outage probability} \\ \cline{4-5} 
			& \multicolumn{2}{c|}{}                   & Simulation results           & Analytical results         \\ \hline
			5	& \multicolumn{2}{c|}{26}                   & $8.0\times 10^{-6}$    &   $2.3\times 10^{-7}$          \\ \hline
			8	& \multicolumn{2}{c|}{70}                   &  $2.60\times 10^{-5}$    &   $3.76\times 10^{-5}$          \\ \hline
			10	& \multicolumn{2}{c|}{96}                   &  $3.2\times 10^{-5}$    &   $2.00\times 10^{-5}$          \\ \hline
			12	& \multicolumn{2}{c|}{104}                   &  $4.5\times 10^{-5}$    &   $6.01\times 10^{-5}$          \\ \hline
			14	& \multicolumn{2}{c|}{122}                   &  $4.6\times 10^{-5}$   &   $6.51\times 10^{-5}$          \\ \hline
			16	& \multicolumn{2}{c|}{139}                   &   $5\times 10^{-5}$  &   $7.5\times 10^{-5}$          \\ \hline
			18	& \multicolumn{2}{c|}{156}                    &   $8.2\times 10^{-5}$   &   $8.00\times 10^{-5}$          \\ \hline
		\end{tabular}
		\label{FOV-OPT}
	\end{table}
	%%%%%%%%%%%%%%%%%%%%%%%%%%%%%%%%%%%%%%
	
\begin{figure}[t]
		\begin{center}
			\includegraphics[width=3.5in]{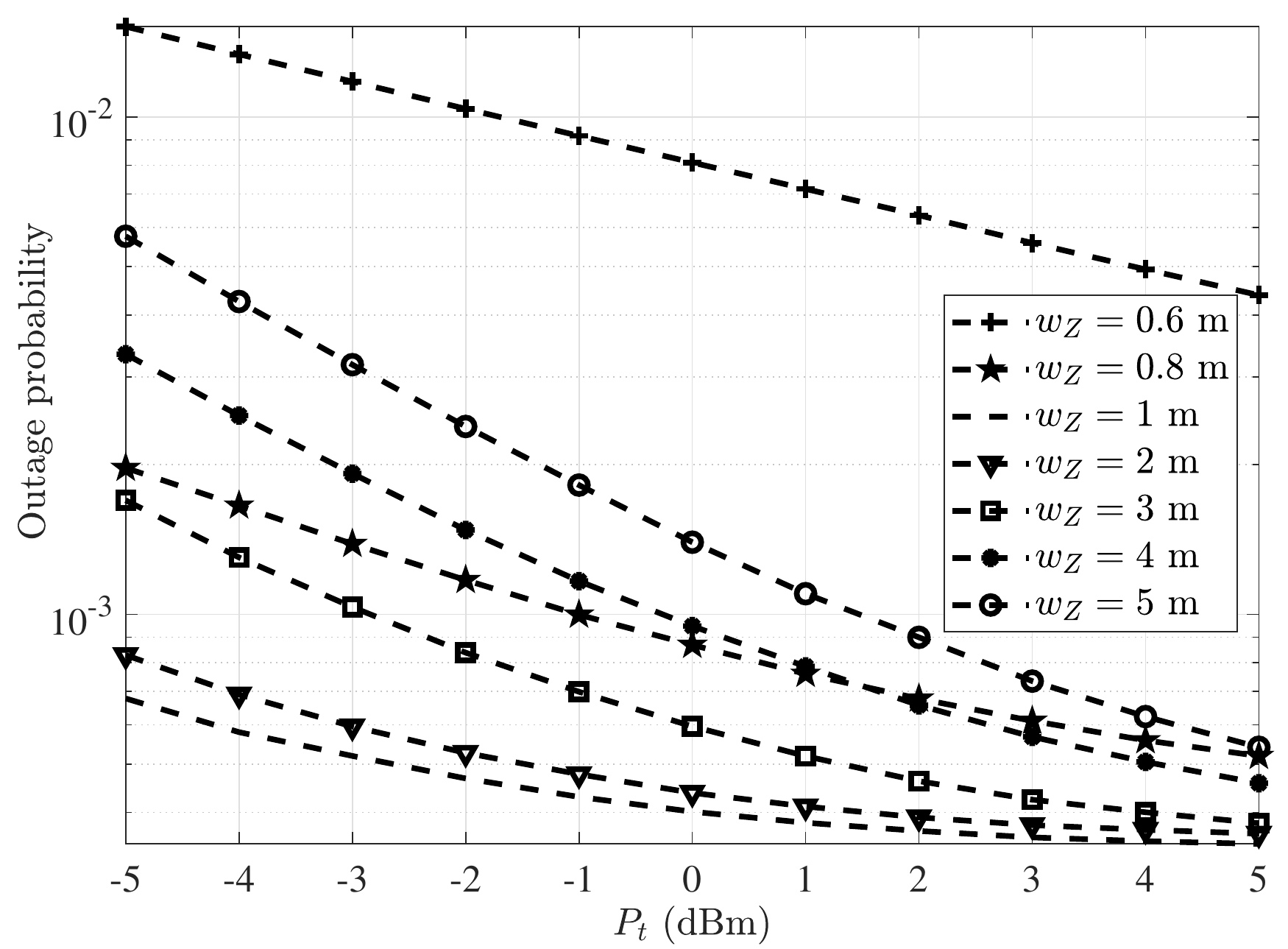}
			\caption{Outage probability versus $P_t$ for $r_a = 5$ cm,  $\sigma_d = 0.4 \textrm{~m}$, $\sigma_0 = 5 \textrm{~mrad}$ and different values of $w_Z$.}
			\label{outage3dadd}
		\end{center}
		%\vspace*{-6pt}
	\end{figure}

\begin{figure}[t]
		\begin{center}
			\includegraphics[width=3.5in]{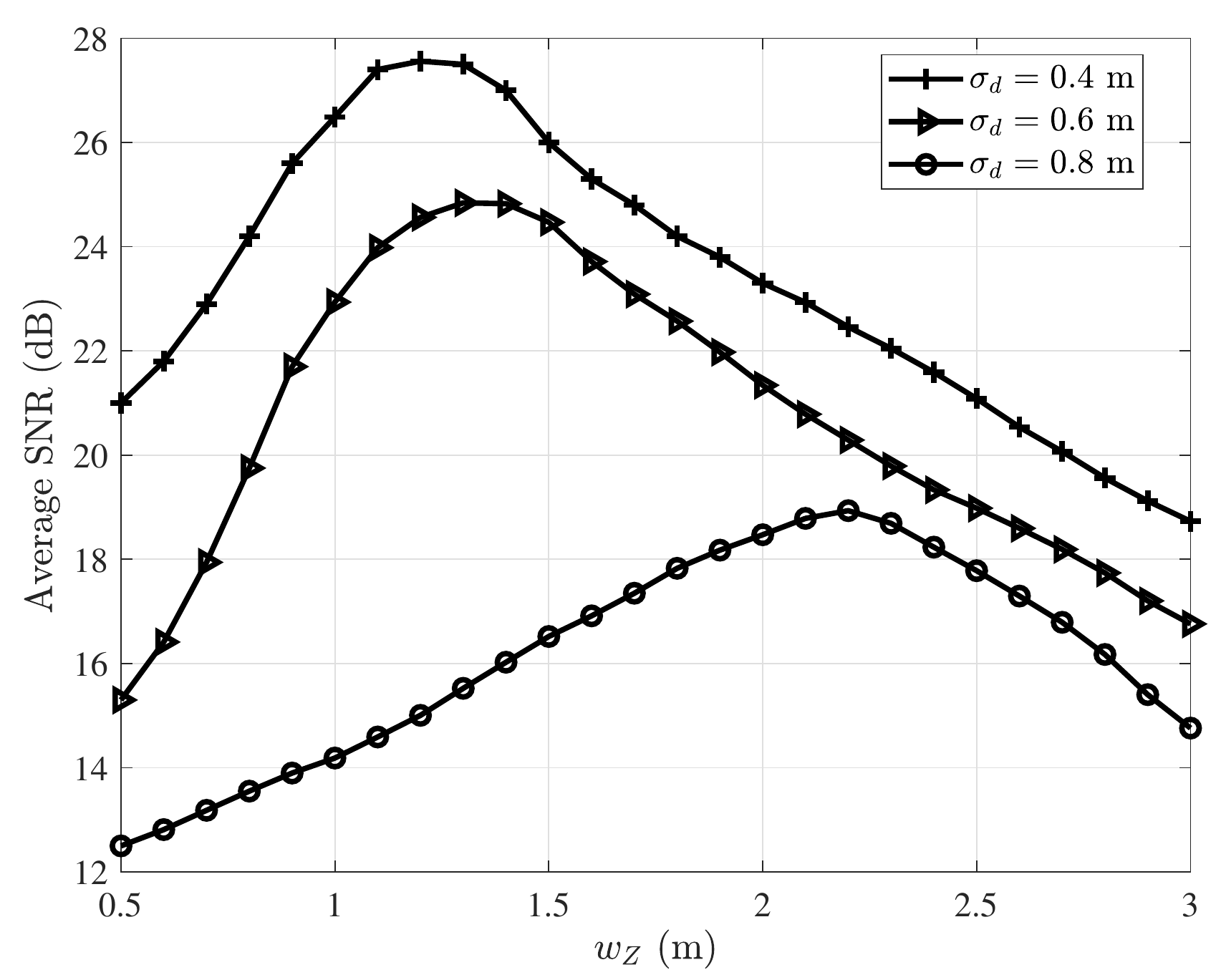}
			\caption{\textcolor{black}{Average SNR versus $w_z$ for $r_a = 10$ cm,   $\sigma_0 = 15 \textrm{~mrad}$, $\theta_{FOV} = $ 75 mrad,  $ Z = 15$ km,  and different values of $\sigma_d$}.}
			\label{SNR_curves}
		\end{center}
		%\vspace*{-6pt}
	\end{figure}

	Figure \ref{outage3dadd} investigates the performance of the considered link by presenting outage probability as the values of transmit power $P_t$ and the beam-width of the transmitter $w_Z$ vary. From this figure, for a given variance of pointing error, it is clear that  the outage performance  largely depends on the values of $w_Z$, and increasing $w_Z$ does not necessarily decrease the amount of the outage probability. Indeed, the effect of $w_Z$ on the outage probability is significant at low transmit power $P_t$. 
	Fig. \ref{outage3dadd}  highlights the need to optimize  $w_Z$ for improving system performance by alleviating the impacts of increasing irradiance fluctuations and beam wander on the performance of the considered link. \textcolor{black}{Moreover, to have more intuition for  link designing and carrying out some tests on the considered link, we have provided the averge SNR curves versus $w_Z$ for different values of instantaneous position fluctuations of the HAP, $\sigma_d$, in Fig. \ref{SNR_curves}. Again, the results of this figure clearly show that choosing the optimal values of $w_Z$ can considerably help  mitigate the effect of receiver vibrations on the link performance. First, it can be seen from  Fig. \ref{SNR_curves} that for each values of $\sigma_d$ there exists an optimal value of $w_Z$ with which we can maximize the average SNR at the receiver. Second, as expected, by increasing $\sigma_d$ the optimal value of $w_Z$ increases to compensate the effect of the AOA fluctuations due to the receiver vibrations on the link performance. Our analytical analysis makes it easy to find the optimal value of  $w_Z$ under different link conditions and  facilitates  the design of ground-to-HAP FSO links  without resorting to time-consuming simulations. }  
	%Moreover, a floor for the outage probability can be noticed at high values of $P_t$ when $h_\textrm{th}$ approaches zero. This error floor can also be realized from the analytical expressions of \eqref{outage_expression} and \eqref{outage_expression_1}. Accordingly, when $h_\textrm{th}$ approaches zero, the amount of outage probability only depends on $\theta_\textrm{FOV}$, $\sigma_0^2$, $\sigma_r^2$, and $\frac{w_z}{r_a}$, and it becomes independent of $P_t$.
	%%%%%%%%%%%%%%%%%%%%%%%%%%%%%%%%%%%%%%
	%%%%%%%%%%%%%%%%
	%%%%%%%%%%%%%%%%
	%%%%%%%%%%%%%%%%%%%%%%%%%%%%%%%%%%%%%%%
	%%%%%%%%%%%%%%%%%%%%%%%%%%%%%%%%%%%%%%
	%%%%%%%%%%%%%%%%
	%%%%%%%%%%%%%%%%
	%%%%%%%%%%%%%%%%%%%%%%%%%%%%%%%%%%%%%%%
	Accordingly, the optimal $w_Z$ values and
	the corresponding outage probabilities obtained from both analytical and simulation results are shown in Table \ref{w_Z-OPT} for different values of
	orientation deviations.  These results  can be applied to find the optimal values of $w_Z$ for different amount of orientation deviations.  
	%%%%%%%%%%%%%%%%%%%%%%%%%%%%%%%%%%%
	\begin{table}
		\def\tablename{Table}
		\centering
		\caption{Optimal values of $W_z$ to achieve minimum outage probability over ground-to-HAP link for different values of $\sigma_d$.}
		%\resizebox{.85\hsize}{!}{%
		\begin{tabular}{|c|c|c|c|c|}
			\hline
			& \multicolumn{4}{c|}{$P_t = 5$ dBm, $\theta_{FOV}$ = 45 mrad, and $\sigma_0$ = 10 mrad}                                          \\ \hline
			\multirow{2}{*}{$\sigma_{d}$ (m)} & \multicolumn{2}{c|}{\multirow{2}{*}{$w_Z$ (m)} } & \multicolumn{2}{c|}{Outage probability} \\ \cline{4-5} 
			& \multicolumn{2}{c|}{}                   & Simulation results           & Analytical results         \\ \hline
			0.1	& \multicolumn{2}{c|}{0.25}                   & $3.90\times 10^{-5}$    &   $1.13\times 10^{-5}$          \\ \hline
			0.2	& \multicolumn{2}{c|}{0.49}                   &  $4.8\times 10^{-5}$    &   $2.84\times 10^{-5}$          \\ \hline
			0.3	& \multicolumn{2}{c|}{0.74}                   &  $4.9\times 10^{-5}$    &   $4.04\times 10^{-5}$          \\ \hline
			0.4	& \multicolumn{2}{c|}{1.10}                   &  $5.3\times 10^{-5}$    &   $4.06\times 10^{-5}$          \\ \hline
			0.6	& \multicolumn{2}{c|}{1.35}                   &  $1.86\times 10^{-4}$   &   $1.03\times 10^{-4}$          \\ \hline
			0.8	& \multicolumn{2}{c|}{2.20}                   &   $2.42\times 10^{-4}$  &   $1.2\times 10^{-4}$          \\ \hline
			1	& \multicolumn{2}{c|}{2.8}                    &   $4.41\times 10^{-4}$   &   $3.10\times 10^{-4}$          \\ \hline
		\end{tabular}
		\label{w_Z-OPT}
	\end{table}
	%%%%%%%%%%%%%%%%%%%%%%%%%%%%%%%%%%%%%%
	
\begin{figure}[t]
		\begin{center}
			\includegraphics[width=3.5 in]{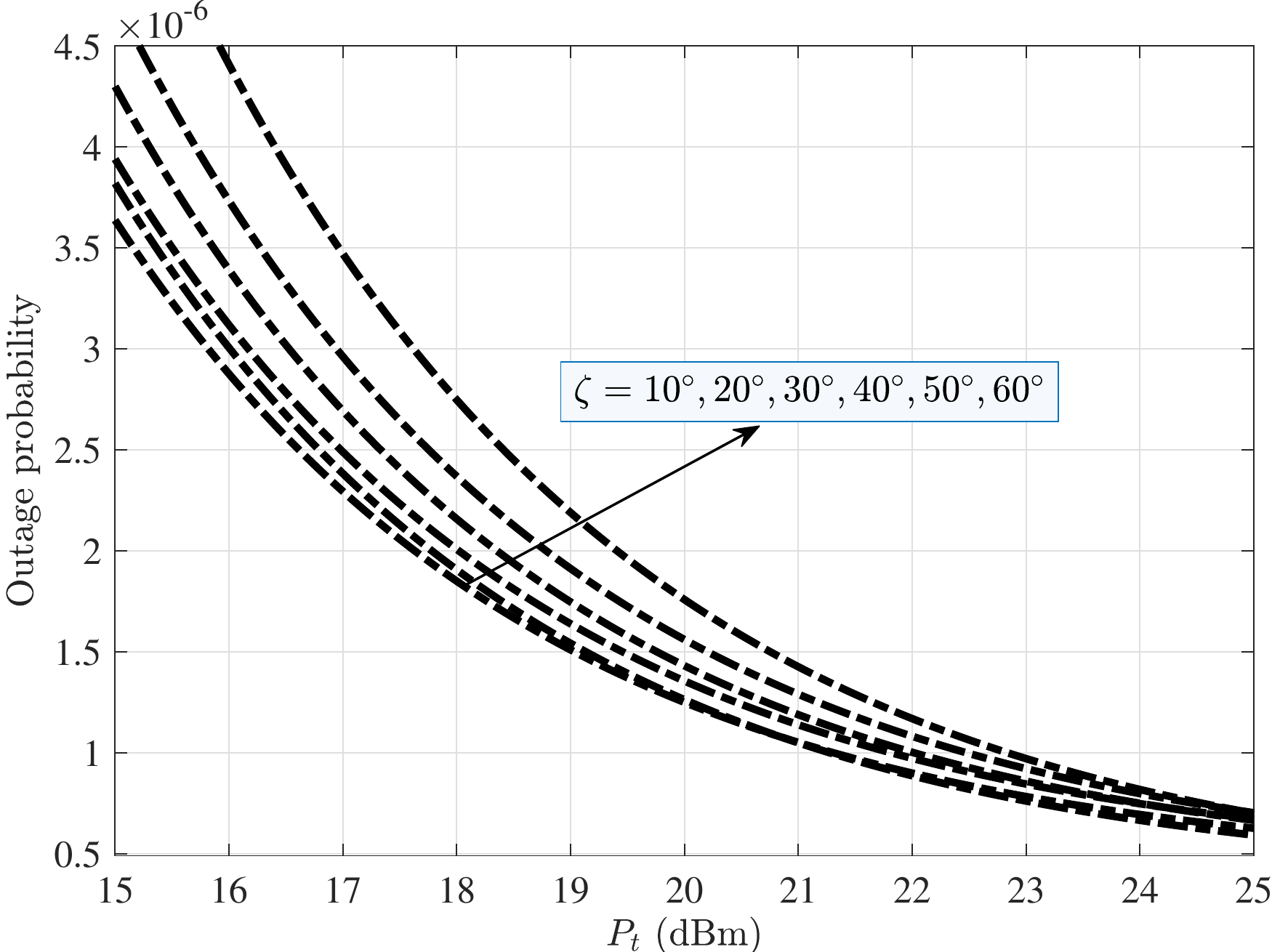}
			\caption{Outage probability versus $P_t$ for  $\sigma_0^2 = $ 10 mrad, $H$ = 17 km, and different values of the zenith angle  $\zeta$ (for each $\zeta$, $w_Z$ is chosen optimally to achieve minimum outage probability).}%Illustration of a typical MR UAV-based FSO communication link.}
			\label{outage_pt_zenith}
		\end{center}
		%\vspace*{-6pt}
	\end{figure}

	Finally,  by using the optimal values of $w_Z$ and $\theta_{\textrm{FOV}}$ obtained from Tables \ref{FOV-OPT} and \ref{w_Z-OPT}, we study in Fig. \ref{outage_pt_zenith} the effect of increasing irradiance fluctuations and beam wander caused by increasing $\zeta$ (or equally increasing the link length $Z$) on link budget. From  Fig. \ref{outage_pt_zenith} and under a given degree of instability of the HAP, we can conclude  that the same system performance for different link lengths can be delivered by employing optimal values for $w_Z$ and $\theta_{\textrm{FOV}}$ and also bearing the cost of increasing the link budget. For instance, to have a same value in outage probability, the change in $\zeta$ from $10^\circ$ to $60^\circ$ will increase transmit power by 2 dBm.
	%%%%%%%%%%%%%%%%%%%%
	%%%%%%%%%%%%%%%%
	%%%%%%%%%%%%%%%%%
	%\begin{figure}
	%%	\vspace*{-0.4cm}
	%	\begin{center}
	%		\subfloat[]{
	%			\includegraphics[width=0.9\linewidth, draft=false]{figures/3D_div_fov_outage_surf5_sigmao5.eps}
	%			\label{fig_4_a}}
	%		\quad
	%		\subfloat[]{
	%			\includegraphics[width=0.9\linewidth, draft=false]{figures/3D_div_fov_outage_surf5_sigmao10.eps}
	%			\label{fig_4_b}}
	%		
	%		\caption{Outage probability versus $\theta_\textrm{div}$ and $\theta_{\textrm{FOV}}$ for $r_a = $ 10 cm, $Z = 20$ km, $P_t$ = 20 dBm, and a) $\sigma_0^2 = $ 5 mrad, and b) $\sigma_0^2 = $ 10 mrad.}
	%	\end{center}
	%	\vspace*{-0.4cm}
	%\end{figure}
	%
	\section{CONCLUSION}
	%%%%%%%%%%%%%%%%%%%%%%%%%%%%%%%%%%%%%%
	We studied the problem of integrating
	FSO communication into HAPs for providing high data rate wireless connectivity. To facilitate design of a ground-to-HAP FSO link, we proposed simple and tractable statistical models for channel model under both LN and GG atmospheric turbulence considerations. The proposed models incorporate the combined effects of position vibrations of the receiver, pointing jitter variance caused by beam wander, detector aperture size, received optical beam-width, the FOV of the receiver, atmospheric attenuation and turbulence, and the AOA fluctuations of the received optical beam. Subsequently, closed-form expressions are derived for the outage probability of the considered link under different turbulence regimes. The developed models make it possible to conduct detailed analysis towards optimizing the transmitted laser beam and the FOV of the receiver in terms of achieving minimum outage probability under different channel conditions. We showed that the performance of such links largely depends
	on the receiver FOV and the received optical beam-width $\omega_Z$. Thus, the results of this paper can be used  for finding the optimal values of link parameters and designing ground-to-HAP FSO links without resorting to time-consuming simulations. 

	\bibliographystyle{IEEEtran}
	\balance
	%\bibliography{IEEEabrv,myref}

\begin{thebibliography}{1}
		\bibitem{mozaffari2019tutorial}
		M.~Mozaffari, W.~Saad, M.~Bennis, Y.~H. Nam, and M.~Debbah, ``A tutorial on
		{UAVs} for wireless networks: Applications, challenges, and open problems,''
		\emph{IEEE Commun. Surveys Tuts.}, Mar. 2019.
		
		\bibitem{karapantazis2005broadband}
		S.~Karapantazis and F.~Pavlidou, ``Broadband communications via high-altitude
		platforms: A survey,'' \emph{IEEE Commun. Surveys Tuts.}, vol.~7, no.~1, pp.
		2--31, May 2005.
		
		\bibitem{mozaffari2017mobile}
		M.~Mozaffari, W.~Saad, M.~Bennis, and M.~Debbah, ``Mobile unmanned aerial
		vehicles {(UAVs)} for energy-efficient internet of things communications,''
		\emph{IEEE Trans. Wireless Commun.}, vol.~16, no.~11, pp. 7574--7589, Nov.
		2017.
		
		\bibitem{alzenad2018fso}
		M.~Alzenad, M.~Z. Shakir, H.~Yanikomeroglu, and M.~S. Alouini, ``{FSO-based
			vertical backhaul/fronthaul framework for 5G+ wireless networks},''
		\emph{IEEE Commun. Mag.}, vol.~56, no.~1, pp. 218--224, Jan. 2018.
		
		\bibitem{dong2018edge}
		Y.~Dong, M.~Z. Hassan, J.~Cheng, M.~J. Hossain, and V.~C. Leung, ``An edge
		computing empowered radio access network with {UAV}-mounted {FSO} fronthaul
		and backhaul: Key challenges and approaches,'' \emph{IEEE Wireless Commun.},
		vol.~25, no.~3, pp. 154--160, June 2018.
		
		\bibitem{loonproject}
		{Google,``Project Loon," [online] https://x.company/loon/}.
		
		\bibitem{faceebook}
		{Facebook,``Internet.org by Facebook," [online] https://internet.org/}.
		
		\bibitem{5}
		C.~M. Schieler \emph{et~al.}, ``{NASA's Terabyte Infrared Delivery (TBIRD)
			Program: Large-Volume Data Transfer from LEO},'' \emph{33rd Annual AIAA/USU
			Conference on Small Satellites}, pp. SSC19--VI--02, 2020.
		
		\bibitem{6}
		E.~Y. Luzhansky \emph{et~al.}, ``{Laser communications relay
			demonstration}.''\hskip 1em plus 0.5em minus 0.4em\relax \emph{SPIE}, vol.
		9739, pp. 102 -- 113, Mar. 2016.
		
		\bibitem{safi2019spatial}
		H.~Safi, A.~Dargahi, and J.~Cheng, ``Spatial beam tracking and data detection
		for an {FSO} link to a {UAV} in the presence of hovering fluctuations,''
		\emph{arXiv preprint arXiv:1904.03774}, 2019.
		
		\bibitem{khan2017gps}
		M.~Khan, M.~Yuksel, and G.~Winkelmaier, ``{GPS-free maintenance of a
			free-space-optical link between two autonomous mobiles},'' \emph{IEEE Trans.
			Mobile Comput.}, vol.~16, no.~6, pp. 1644--1657, June 2017.
		
		\bibitem{kaadan2016spherical}
		A.~Kaadan, H.~Refai, and P.~Lopresti, ``{Spherical FSO receivers for UAV
			communication: Geometric coverage models},'' \emph{IEEE Trans. Aerosp.
			Electron. Syst.}, vol.~52, no.~5, pp. 2157--2167, Oct. 2016.
		
		\bibitem{kaymak2018survey}
		Y.~Kaymak, R.~Rojas-Cessa, J.~Feng, N.~Ansari, M.~Zhou, and T.~Zhang, ``A
		survey on acquisition, tracking, and pointing mechanisms for mobile
		free-space optical communications,'' \emph{IEEE Commun. Surveys Tuts.},
		vol.~20, no.~2, pp. 1104--1123, Feb. 2018.
		
		\bibitem{laserbook}
		L.~C. Andrews and R.~L. Phillips, \emph{Laser Beam Propagation through Random
			Media}.\hskip 1em plus 0.5em minus 0.4em\relax SPIE Press Bellingham, WA,
		2005.
		
		\bibitem{song2014robust}
		T.~Song and P.-Y. Kam, ``{A robust GLRT receiver with implicit channel
			estimation and automatic threshold adjustment for the free space optical
			channel with IM/DD},'' \emph{J. Lightw. Technol.}, vol.~32, no.~3, pp.
		369--383, Feb. 2014.
		
		\bibitem{dabiri2017glrt}
		M.~T. Dabiri, S.~M.~S. Sadough, and H.~Safi, ``{GLRT-based sequence detection
			of OOK modulation over FSO turbulence channels},'' \emph{IEEE Photon.
			Technol. Lett.}, vol.~29, no.~17, pp. 1494--1497, Jan. 2017.
		
		\bibitem{jaiswal2018investigation}
		A.~Jaiswal, M.~Abaza, M.~R. Bhatnagar, and V.~K. Jain, ``An investigation of
		performance and diversity property of optical space shift keying based
		{FSO-MIMO} system,'' \emph{IEEE Trans. Commun.}, vol.~66, no.~9, pp.
		4028--4042, Mar. 2018.
		
		\bibitem{mendenhall2007design}
		J.~A. Mendenhall \emph{et~al.}, ``Design of an optical photon counting array
		receiver system for deep-space communications,'' \emph{in Proceedings of the
			IEEE}, vol.~95, no.~10, pp. 2059--2069, Oct. 2007.
		
		\bibitem{10.1117/12.273685}
		I.~I. Kim, H.~Hakakha, P.~Adhikari, E.~J. Korevaar, and A.~K. Majumdar,
		``{Scintillation reduction using multiple transmitters}.''\hskip 1em plus
		0.5em minus 0.4em\relax \emph{SPIE}, vol. 2990, pp. 102 -- 113, Apr. 1997.
		
		\bibitem{safi2019adaptive}
		H.~Safi, A.~A. Sharifi, M.~T. Dabiri, I.~S. Ansari, and J.~Cheng, ``Adaptive
		channel coding and power control for practical {FSO} communication systems
		under channel estimation error,'' \emph{IEEE Trans. Vehic. Technol.},
		vol.~68, no.~8, pp. 7566--7577, Aug. 2019.
		
		\bibitem{walther2010air}
		F.~G. Walther, S.~Michael, R.~R. Parenti, and J.~A. Taylor, ``Air-to-ground
		lasercom system demonstration design overview and results summary,'' \emph{in
			Free Space Laser Communication, International Society for Optics and
			Photonics}, vol. 7814, p. 78140Y, Aug. 2010.
		
		\bibitem{AOT}
		R.~K. Tyson, ``Bit error rate for free space adaptive optics laser
		communications,'' \emph{J. Opt. Soc. Amer. A, Opt. Image Sci.}, vol.~19,
		no.~4, pp. 753--758, Apr. 2002.
		
		\bibitem{AOT1}
		A.~K. Majumdar and J.~C. Riclkin, \emph{Free-Space Laser Communications:
			Principles And Advances}.\hskip 1em plus 0.5em minus 0.4em\relax
		Springer-Verlag, New York, NY, USA, 2007.
		
		\bibitem{AOT2}
		H.~Hemmati, \emph{Deep Space Optical Communications}.\hskip 1em plus 0.5em
		minus 0.4em\relax Wiley-Interscience, Hoboken, NJ, USA, 2006.
		
		\bibitem{khalighi2014}
		M.~A. Khalighi and M.~Uysal, ``Survey on free space optical communication:{ A
			communication theory perspective},'' \emph{IEEE Commun. Surveys Tuts.},
		vol.~16, no.~4, pp. 2231--2258, Nov. 2014.
		
		\bibitem{kaushal2017optical}
		H.~Kaushal and G.~Kaddoum, ``{Optical communication in space: Challenges and
			mitigation techniques},'' \emph{IEEE Commun. Surveys Tuts.}, vol.~19, no.~1,
		pp. 57--96, Aug. 2017.
		
		\bibitem{AOT3}
		Stotts, Larry \textit{et al.}, ``Free-space optical
		communications link budget estimation,'' \emph{Appl. Opt.}, vol.~49, no.~28,
		pp. 5333--5343, Oct. 2010.
		
		\bibitem{huang2017free}
		S.~Huang and M.~Safari, ``Free-space optical communication impaired by angular
		fluctuations,'' \emph{IEEE Trans. Wireless Commun.}, vol.~16, no.~11, pp.
		7475--7487, Nov. 2017.
		
		\bibitem{fawaz2018uav}
		W.~Fawaz, C.~Abou-Rjeily, and C.~Assi, ``{UAV-aided cooperation for FSO
			communication systems},'' \emph{IEEE Commun. Mag.}, vol.~56, no.~1, pp.
		70--75, Jan. 2018.
		
		\bibitem{yang2019performance}
		L.~Yang, J.~Yuan, X.~Liu, and M.~O. Hasna, ``On the performance of {LAP-Based
			Multiple-Hop RF/FSO} systems,'' \emph{IEEE Trans. Aerosp. Electron. Syst},
		vol.~55, no.~1, pp. 499--505, Feb. 2019.
		
		\bibitem{li201780}
		L.~Li, R.~Zhang, Z.~Zhao, G.~Xie, P.~Liao, K.~Pang, H.~Song, C.~Liu, Y.~Ren,
		G.~Labroille \emph{et~al.}, ``{80-Gbit/s 100-m free-space optical data
			transmission link via a flying UAV using multiplexing of
			orbital-angular-momentum beams},'' \emph{arXiv preprint arXiv:1708.02923},
		2017.
		
		\bibitem{mai2019beam}
		V.~Mai and H.~Kim, ``Beam size optimization and adaptation for high-altitude
		airborne free-space optical communication systems,'' \emph{IEEE Photon. J.},
		vol.~11, no.~2, pp. 1--13, Feb. 2019.
		
		\bibitem{li2018investigation}
		M.~Li, Y.~Hong, C.~Zeng, Y.~Song, and X.~Zhang, ``Investigation on the
		{UAV}-to-satellite optical communication systems,'' \emph{IEEE J. Sel. Areas
			Commun.}, vol.~36, no.~9, pp. 2128--2138, Sep. 2018.
		
		\bibitem{dabiri2018channel}
		M.~T. Dabiri, S.~M.~S. Sadough, and M.~A. Khalighi, ``Channel modeling and
		parameter optimization for hovering {UAV-based} free-space optical links,''
		\emph{IEEE J. Sel. Areas Commun.}, vol.~36, no.~9, pp. 2104--2113, Sep. 2018.
		
		\bibitem{dabiri2019tractable}
		M.~T. Dabiri, S.~M.~S. Sadough, and I.~S. Ansari, ``Tractable optical channel
		modeling between {UAVs},'' \emph{IEEE Trans. Vehic. Technol.}, vol.~68, no.~12, pp. 11543--11550, Sep. 2019.
		
		\bibitem{najafi2019statistical}
		M.~Najafi, H.~Ajam, V.~Jamali, P.~D. Diamantoulakis, G.~K. Karagiannidis, and
		R.~Schober, ``Statistical modeling of the fso fronthaul channel for uav-based
		networks,'' \emph{arXiv preprint arXiv:1905.12424}, 2019.
		
		\bibitem{gagliardi1995optical}
		R.~M. Gagliardi and S.~Karp, \emph{Optical Communications}.\hskip 1em plus
		0.5em minus 0.4em\relax New York, Wiley-Interscience, 1995.
		
		\bibitem{zhu2002free}
		X.~Zhu and J.~M. Kahn, ``{Free-space optical communication through atmospheric
			turbulence channels},'' \emph{IEEE Trans. Commun.}, vol.~50, no.~8, pp.
		1293--1300, Aug. 2002.
		
		\bibitem{dabiri2017fso}
		M.~T. Dabiri, S.~M.~S. Sadough, and M.~A. Khalighi, ``{FSO channel estimation
			for OOK modulation with APD receiver over atmospheric turbulence and pointing
			errors},'' \emph{Opt. Commun.}, vol. 402, pp. 577--584, Nov. 2017.
		
		\bibitem{NB2}
		A.~Carrasco-Casado, J.~M. S{\'a}nchez-Pena, and R.~Vergaz, ``{CTA} telescopes
		as deep-space lasercom ground receivers,'' \emph{IEEE Photon. Journal.},
		vol.~12, no.~4, pp. 1--4, Nov. 2015.
		
		\bibitem{NB3}
		S.~Riechelmann, M.~Schrempf, and G.~Seckmeyer, ``Simultaneous measurement of
		spectral sky radiance by a non-scanning multidirectional spectroradiometer
		{(MUDIS)},'' \emph{Measurement Science and Technology}, vol.~24, no.~12, p.
		125501, Nov. 2013.
		
		\bibitem{NB4}
		E.~E. Bell, L.~Eisner, J.~Young, and R.~A. Oetjen, ``Spectral radiance of sky
		and terrain at wavelengths between 1 and 20 microns. ii. sky measurements,''
		\emph{JOSA}, vol.~50, no.~12, pp. 1313--1320, Dec. 1960.
		
		\bibitem{NB1}
		H.~Hemmati, \emph{Deep Space Optical Communications}.\hskip 1em plus 0.5em
		minus 0.4em\relax Wiley-Interscience,, Hoboken, NJ, USA, 2006.
		
		\bibitem{pointing2007}
		A.~A. Farid and S.~Hranilovic, ``Outage capacity optimization for free-space
		optical links with pointing errors,'' \emph{IEEE/OSA J. Light. Technol.},
		vol.~25, no.~7, pp. 1702--1710, July 2007.
		
		\bibitem{al2004fog}
		M.~C. Al~Naboulsi, H.~Sizun, and F.~de~Fornel, ``Fog attenuation prediction for
		optical and infrared waves,'' \emph{Opt. Eng.}, vol.~43, no.~2, pp. 319--330,
		Feb. 2004.
		
		\bibitem{H-V}
		R.~E. Hufnagel, \emph{Variations of Atmospheric Turbulence}.\hskip 1em plus
		0.5em minus 0.4em\relax Tech. Report, 1974.
		
		\bibitem{H-V2}
		R.~K. Tyson, ``Adaptive optics and ground-to-space laser communication,''
		\emph{Appl. Opt.}, vol.~35, no.~19, pp. 3640--3646, July 1996.
		
		\bibitem{H-V3}
		G.~C. Valley, ``Isoplanatic degradation of tilt correction and short-term
		imaging systems,'' \emph{Appl. Opt.}, vol.~19, no.~4, pp. 574--577, Feb.
		1980.
		
		\bibitem{saleh2019fundamentals}
		B.~E. Saleh and M.~C. Teich, \emph{Fundamentals of Photonics}.\hskip 1em plus
		0.5em minus 0.4em\relax John Wiley \& Sons, 2019.
		
		\bibitem{siegman1986lasers}
		A.~E. Siegman, \emph{Lasers University Science Books}.\hskip 1em plus 0.5em
		minus 0.4em\relax Mill Valley, CA, 1986.
		
		\bibitem{orsag2017dexterous}
		M.~Orsag, C.~Korpela, S.~Bogdan, and P.~Oh, ``Dexterous aerial robots—mobile
		manipulation using unmanned aerial systems,'' \emph{IEEE Trans. Robot.},
		vol.~33, no.~6, pp. 1453--1466, Dec. 2017.
		
		\bibitem{lee2017estimation}
		H.~Lee and H.~J. Kim, ``Estimation, control, and planning for autonomous aerial
		transportation,'' \emph{IEEE Trans. Indus. Elec.}, vol.~64, no.~4, pp.
		3369--3379, Aug. 2017.
		
		\bibitem{faessler2017thrust}
		M.~Faessler, D.~Falanga, and D.~Scaramuzza, ``Thrust mixing, saturation, and
		body-rate control for accurate aggressive quadrotor flight,'' \emph{IEEE
			Robot. Automat. Lett.}, vol.~2, no.~2, pp. 476--482, Dec. 2017.
		
		\bibitem{born2013principles}
		M.~Born and E.~Wolf, \emph{Principles of Optics: Electromagnetic Theory of
			Propagation, Interference and Diffraction of Light}.\hskip 1em plus 0.5em
		minus 0.4em\relax Cambridge Univ. Press, 1999.
		
		\bibitem{ghassemlooy2012optical}
		Z.~Ghassemlooy, W.~Popoola, and S.~Rajbhandari, \emph{Optical Wireless
			Communications}.\hskip 1em plus 0.5em minus 0.4em\relax CRC Press Boca Raton,
		FL, 2012.
	\end{thebibliography}

\end{document}